\documentclass[prd]{revtex4}
\usepackage{graphics}
\newcommand\half{\frac{1}{2}}
\begin{document}
\title{Uncertainty in the leading order PQCD calculations of 
$B$ meson decays}
\author{Takeshi Kurimoto }
\email{krmt@sci.u-toyama.ac.jp}
\affiliation{%
Faculty of Science, University of Toyama, 
Toyama 930-8555, Japan}
\begin{abstract}
Uncertainty in the PQCD calculation of $B$ decays is investigated 
in $B\rightarrow \pi$, $B\rightarrow D$ transition form factors 
and $B\rightarrow D\pi$ decay amplitudes. $B$ meson distribution 
amplitude dependence is studied by taking three kinds of distribution 
amplitudes so far suggested. It is found that almost same $q^2$ dependence 
of the form factors can be obtained irrespective of the types of 
the $B$ meson distribution amplitudes by suitably choosing one parameter. 
$B\rightarrow D\pi$ process shows the difference due to the 
distribution amplitude.
The effect of the sub-leading component of the $B$ meson distribution 
amplitude is also studied in the three processes. The numerical 
results of calculations with the sub-leading component 
can be well approximated by the 
leading order calculation with a suitable choice of the 
distribution amplitude parameters.    
\end{abstract}
\maketitle
\section{Introduction}
$B$ meson decays has been attracting much attention to check 
the consistency of the standard model (SM) and to explore the existence 
of a new physics beyond the SM.  Two $B$ physics dedicated experimental 
facilities are constructed at KEK and SLAC.
The Belle and the BABAR groups have reported 
a lot of interesting results since their beginnings\cite{BFACS}. 
Many fruitful theoretical works on $B$ physics have been 
made in these decades, but hadronic effects often obscure the 
theoretical predictions. Li and collaborators 
developed the so-called PQCD method and applied it
to exclusive $B$ meson decays as one of the approaches to tackle 
this issue\cite{LiPQCD}. The PQCD method gives  
reasonable predictions on $B\rightarrow K\pi$\cite{KLS}, 
$B\rightarrow \pi\pi$\cite{LUY} and other $B$ decays\cite{PQCDB}.

In PQCD method, a decay amplitude is obtained as a convolution 
of a hard part ($H$) and meson distribution amplitudes ($\phi_k$).
\begin{equation}
 Amp = \int \phi_1 \times H \times \phi_2\cdots\; .
\end{equation} 
The hard part can in principle be perturbatively calculated in a systematic way, 
while the non-perturbative contributions are incorporated 
into the distribution amplitudes. 
Major uncertainty in the PQCD calculation lies in the choice of
distribution amplitudes.
We need a model or a non-perturbative method
like QCD based sum-rule\cite{PB2,PB1,PB4,BB} to obtain 
the distribution amplitudes.
Meson distribution amplitudes are important
also in the study of $B$ non-leptonic decays with
QCD factorization method\cite{BBNS} and
in the calculation of the form factors with 
QCD based sum-rule scheme\cite{PB1,PB4}.
So far most of the PQCD calculations of $B$ decays are given
in the leading order of $\alpha_S$ and $1/M$. 
(A trial to estimate the higher order effects
in $\alpha_S$ is given in \cite{MSSA}.)
The aim of this paper is to investigate the uncertainty of 
the leading order PQCD calculations. 
Our strategy is as follows: In Sec.II, we analyze $B\rightarrow\pi$ 
form factors to estimate the uncertainty due to the factors 
given below; 
\begin{enumerate}
\item  $B$ meson distribution amplitude: 
$B\rightarrow\pi$ form factors are calculated by adopting
three kinds of $B$ meson distribution amplitudes proposed in the 
previous works\cite{GN,KLS,KKQT}. 
The parameters of $B$ meson distribution amplitudes are fixed to 
accommodate with the reasonable value of the form factor at $q^2=0$. 
Then we vary these parameters to see how the value of the form factor changes.
\item Pion distribution amplitude: We adopt the distribution amplitude
given in QCD based sum-rule analysis\cite{PB2,PB1}, and  
investigate the dependence on the parameters of those 
distribution amplitudes.
(The effect of choosing another pion distribution amplitude is 
investigated in \cite{HuWu}.)
\item Hard part: The dependence on $\Lambda_{QCD}$ and other renormalization 
group parameters are investigated.
\item Sub-leading contributions: We estimate the $O(1/M)$ corrections in the hard part 
and the contributions from the sub-leading component of the $B$ meson 
distribution amplitude.
\end{enumerate} 
In Sec. III, we analyze $B\rightarrow D$ form factors. The parameters of 
$B$ meson distribution amplitudes are fixed by the first analysis. 
Here, we investigate the dependence on the parameter of 
the $D$ meson distribution amplitude proposed in \cite{TLS2}.
In Sec. IV, we analyze $B\rightarrow D\pi$ decays 
by using the $B$, $D$ and pion distribution 
amplitudes fixed in the previous analyses. 
The non-factorizable contribution is important 
in $B\rightarrow D\pi$ decays \cite{KKLL}.  
We show which $B$ meson distribution amplitude gives better results 
by calculating the non-factorizable contribution.   
Sec. V is devoted to summary and discussions.
\section{Heavy-to-light Form Factors}
We first analyze the $B \rightarrow \pi$ 
form factors in the fast recoil region with PQCD method. 
We shall determine the parameters of 
$B$ meson distribution amplitudes from the $B\to\pi$ form factors. 
The $B\to\pi$ transition form factors $F^{B\pi}_+$ and
$F^{B\pi}_0$ are defined by the matrix element,
\begin{equation}
\langle\pi(P_2)|{\bar b}(0)\gamma_\mu u(0)|B(P_1)\rangle 
=F^{B\pi}_+(q^2)\left[(P_1+P_2)_\mu-\frac{m_B^2-m_\pi^2}{q^2}q_\mu\right]
 +F^{B\pi}_0(q^2)\frac{m_B^2-m_\pi^2}{q^2}q_\mu\;,
\end{equation}
where $q=P_1-P_2$ is the lepton-pair momentum. Another equivalent
definition is
\begin{eqnarray}
\langle\pi(P_2)|{\bar b}(0)\gamma_\mu u(0)|B(P_1)\rangle
=f_1(q^2)P_{1\mu}+f_2(q^2)P_{2\mu}\;,
\end{eqnarray}
in which the form factors $f_1$ and $f_2$ are related to
$F^{B\pi}_+$ and $F^{B\pi}_0$ by
\begin{eqnarray}
F^{B\pi}_+&=&\half(f_1+f_2)\;,
\label{FormBpi1}\\
F^{B\pi}_0&=&\half f_1\left(1+ \frac{q^2}{m_B^2}\right) +\half
f_2\left(1 - \frac{q^2}{m_B^2}\right)\;.
\end{eqnarray}

\begin{figure}[hb]
\resizebox{8cm}{!}{%
\includegraphics{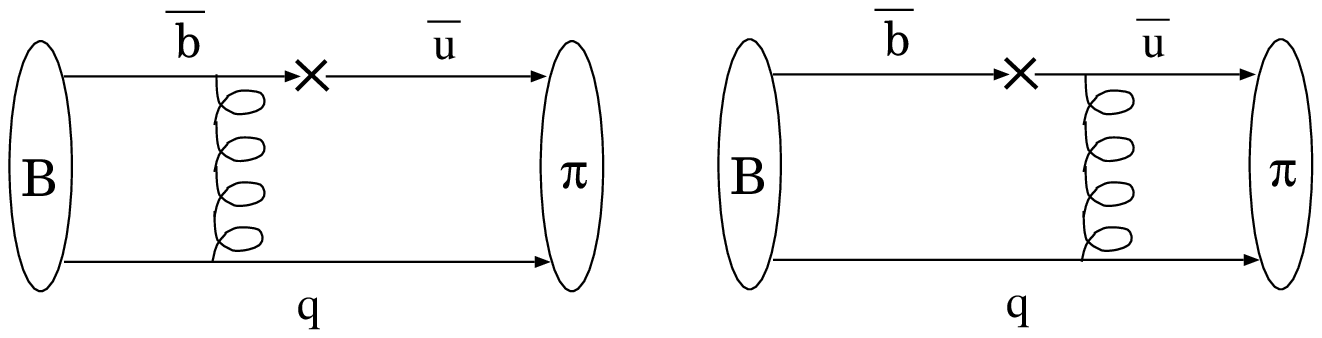}
}
\caption{Leading-order contribution to $F^{B\pi}$.}
\label{Bpiform1}
\end{figure}

In PQCD method, the form factors $F^{B\pi}_{+,0}$ are
derived from the diagrams with one hard gluon exchange shown in
Fig.~\ref{Bpiform1}. PQCD works best in the region
with large energy transfer, i.e., with small $q^2$. Soft
contribution from the diagram without any hard gluon is Sudakov
suppressed\cite{TLS}.
The formulae for the $B\to\pi$ form factors are
given as
\begin{eqnarray}
f_1&=&16\pi m_B^2C_Fr_\pi\int dx_1dx_2\int b_1db_1
b_2db_2\phi_B(x_1,b_1) [\phi_\pi^p(x_2)-\phi_\pi^t(x_2)]
\nonumber\\
&&\times
 E(t^{(1)})h(x_1,x_2,b_1,b_2)\;,
\label{FormBpi2}\\
f_2&=&16\pi m_B^2C_F\int dx_1dx_2\int b_1db_1
b_2db_2\phi_B(x_1,b_1)
\nonumber\\
&\times& \Biggl\{\left[\phi_\pi(x_2)(1+x_2\eta)
+2r_\pi\left((\frac{1}{\eta} -x_2 )\phi_\pi^t(x_2) -x_2\phi_\pi^p(x_2)
 \right)\right]E(t^{(1)})h(x_1,x_2,b_1,b_2)
\nonumber\\
& &+ 2r_\pi\phi_\pi^p E(t^{(2})h(x_2,x_1,b_2,b_1)\Biggr\}\;,
\label{FormBpi3}
\end{eqnarray}
with $\eta = 2P_1\cdot P_2/m_B^2=1-(q^2/m_B^2)$, the ratio
$r_\pi=m_0/m_B$ ($m_0$: chiral mass of pion) and the evolution
factor
\begin{eqnarray}
E(t)=\alpha_s(t)e^{-S_B(t)-S_\pi(t)}\;,
\label{evol}
\end{eqnarray}
where $S_B$ and $S_\pi$ are the Sudakov factor of $k_T$ part
for $B$ meson and pion, respectively\cite{TLS}.
The hard function is given as
\begin{eqnarray}
h(x_1,x_2,b_1,b_2)&=&S_t(x_2)K_{0}\left(\sqrt{x_1x_2\eta}m_Bb_1\right)
\nonumber \\
& &\times \left[\theta(b_1-b_2)K_0\left(\sqrt{x_2\eta}m_B
b_1\right)I_0\left(\sqrt{x_2\eta}m_Bb_2\right)\right.
\nonumber \\
& &\left.+\theta(b_2-b_1)K_0\left(\sqrt{x_2\eta}m_Bb_2\right)
I_0\left(\sqrt{x_2\eta}m_Bb_1\right)\right]\;, \label{dh}
\end{eqnarray}
where the factor $S_t$ is the threshold resummation factor
\begin{eqnarray}
S_t(x)=\frac{2^{1+2c}\Gamma(3/2+c)}{\sqrt{\pi}\Gamma(1+c)} [x(1-x)]^c\;,
\label{trs}
\end{eqnarray}
which suppresses the end-point behaviors of the meson distribution
amplitudes. The hard scales $t^{(1),(2)}$ are defined as
\begin{eqnarray}
t^{(1)}&=&{\rm max}(\sqrt{x_2\eta}m_B,1/b_1,1/b_2)\;,
\nonumber\\
t^{(2)}&=&{\rm max}(\sqrt{x_1\eta}m_B,1/b_1,1/b_2)\;.
\label{scale_t}\end{eqnarray}

\begin{figure}[hb]
\resizebox{6cm}{!}{
\includegraphics{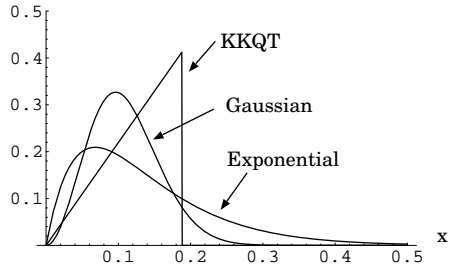}}
\caption{$\phi_B(x,0)$ for $\omega_{KLS}=0.38$, 
$\omega_{GN}=0.36$ and $\Lambda_{KKQT}/M_B=0.094$, }
\label{bwvfig}
\end{figure}

We investigate here the following candidates of $B$ meson distribution 
amplitude:
\begin{eqnarray}
\phi_B^{KLS}(x,b)&=&N_B^{KLS}x^2(1-x)^2
\exp\left[-\frac{1}{2}\left(\frac{xm_B}{\omega_{KLS}}\right)^2
-\frac{\omega_{KLS}^2 b^2}{2}\right]\;, \label{ourBwv}\\
\phi_B^{GN}(x,b)&=&N_B^{GN} x\; \exp\left[-\frac{xm_B}{\omega_{GN}}
\right]\frac{1}{1 + (b\omega_{GN})^2} \;, \label{GNwv}\\ 
\phi_B^{KKQT}(x,b)&=&N_B^{KKQT} x\;
\theta(x)\theta(\frac{2\Lambda_{KKQT}}{m_B} -x)  
J_0 \left(
        b\sqrt{x(\frac{2\Lambda_{KKQT}}{m_B} -x)}\right)\;. \label{KKQTwv}
\end{eqnarray} 
The first one, which we call Gaussian type, is proposed in \cite{KLS}.
The $x$ dependence of the second one, which we call exponential type, 
is proposed in \cite{GN}, and we take its $b$ dependence as Lorentzian, 
the Fourier transform of the exponential function. 
The third one, which we call KKQT type, is obtained by solving the equations 
of motion under the approximation of neglecting 3-parton 
contributions\cite{KKQT}.
Each candidate is parameterized by one parameter, 
$\omega_{KLS}$, $\omega_{GN}$ or $\Lambda_{KKQT}$.
The normalization constant $N_B$ is related to the decay constant
$f_B$ through the relation
\begin{eqnarray}
\int dx\; \phi_B(x,0)=\frac{f_B}{2\sqrt{2N_c}}\;.
\end{eqnarray}
The shapes of these $B$ meson distribution amplitude with $b=0$ are 
shown in Fig. \ref{bwvfig}, where the parameters are chosen so that 
$F^{B\pi}_{+,0}(0) \cong 0.3$ as explained later.

In Eqs.~(\ref{FormBpi2}) and (\ref{FormBpi3}) we have included the
two-parton twist-3 distribution amplitudes $\phi^p_\pi$
and $\phi^t_\pi$ associated with the pseudo-scalar and
pseudo-tensor structures of the pion, respectively\cite{PB2}. The
contribution from the axial vector component\ $\phi_\pi$ is
twist-2. The pion distribution amplitudes derived from QCD based sum
rule are given as\cite{PB1}
\begin{eqnarray}
\phi_\pi(x)&=&\frac{3f_\pi}{\sqrt{2N_c}} x(1-x)
\left[1+ a_2 C_2^{3/2}(1-2x)+ a_4 C_4^{3/2}(1-2x)\right]\;,
\label{pioa}\\
\phi_\pi^p(x)&=&\frac{f_\pi}{2\sqrt{2N_c}}
\left[1+a_{2p} C_2^{1/2}(1-2x)+ a_{4p} C_4^{1/2}(1-2x)\right]\;,
\label{piob}\\
\phi_\pi^t(x)&=&\frac{f_\pi}{2\sqrt{2N_c}} (1-2x)
\left[1+6 a_{2t}(10x^2-10x+1)\right]\;.
\label{pioc}
\end{eqnarray}
The coefficients $a_2\;, \dots , a_{2t}$ are defined as\cite{PB5}
\begin{eqnarray}
a_2(1\text{ GeV}) &=& 0.44\ ,\ a_4(1\text{ GeV}) = 0.25\\
a_{2p} &=& 30 \eta_3  -\frac{5}{2} \rho_\pi^2 \ ,\
a_{4p} = -(3\eta_3\omega_3  +\frac{27}{20} \rho_\pi^2  +
                  \frac{81}{10} \rho_\pi^2 a_2),
\label{gegenap} \\
a_{2t} &=& 5\eta_3 -\frac{1}{2}\eta_3\omega_3  -\frac{7}{20} \rho_\pi^2
                           -  \frac{3}{5} \rho_\pi^2 a_2,
\label{gegenat}
\end{eqnarray}
where
\begin{eqnarray}
\rho_\pi^2 &=& \frac{(m_d + m_u)}{m_0} =
\frac{m_\pi^2}{m_0^2}\ ,\label{defrho}\\
a_k (\mu) &=&  \left[ \frac{\alpha_s(\mu)}{\alpha_s(\mu_0)}
\right]^{\gamma_k/b} a_k(\mu_0)\ ,\
b=11- \frac{2}{3}N_F\ ,
\label{akdep}\\
 \gamma_k &=& 4C_F \left[\psi(k+2) + \gamma_E -\frac{3}{4}
         - \frac{1}{2(k+1)(k+2)}\right]\ ,\\
\eta_3(\mu) &=&  \left[ \frac{\alpha_s(\mu)}{\alpha_s(\mu_0)}
\right]^{\gamma^\eta_3/b} \eta_3 (\mu_0)\ , \
\gamma^\eta_3 = \frac{16}{3} C_F + N_C, \\
\omega_3(\mu) &=&  \left[ \frac{\alpha_s(\mu)}{\alpha_s(\mu_0)}
\right]^{\gamma^\omega_3/b}
 \omega_3 (\mu_0)\ , \
\gamma^\omega_3 = -\frac{25}{6} C_F +\frac{7}{3} N_C\ ,
\label{omegadep}
\end{eqnarray}
with $\eta_3(1\text{ GeV}) = 0.015$, $\omega_3(1\text{ GeV}) = -3$.
The Gegenbauer polynomials are defined by
\begin{eqnarray}
& &C_2^{1/2}(t)=\frac{1}{2}(3t^2-1)\;,\;\;\;
C_4^{1/2}(t)=\frac{1}{8}(35 t^4 -30 t^2 +3)\;,
\nonumber\\
& &C_2^{3/2}(t)=\frac{3}{2}(5t^2-1)\;,\;\;\;
C_4^{3/2}(t)=\frac{15}{8}(21 t^4 -14 t^2 +1) \;.
\end{eqnarray}
\begin{figure}[h]
\resizebox{6cm}{!}{
\includegraphics{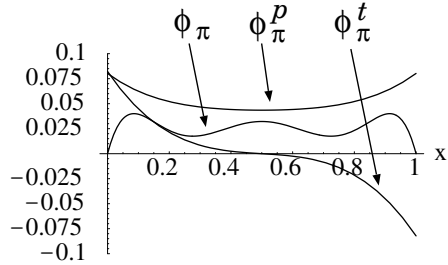}}
\caption{$\phi_\pi(x)$, $\phi^p_\pi(x)$ and $\phi^t_\pi(x)$ for
default values of inputs.}
\label{Piwvfig}
\end{figure}
\subsection{Numerical results}

\begin{figure}[h]
\resizebox{8cm}{!}{\includegraphics{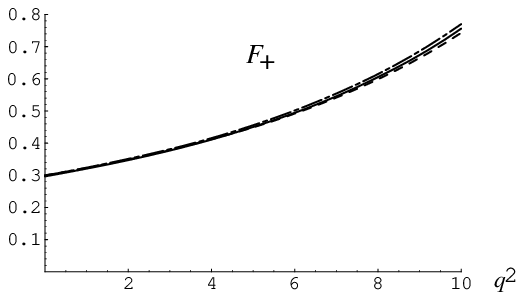}}\ \
\resizebox{7cm}{!}{\includegraphics{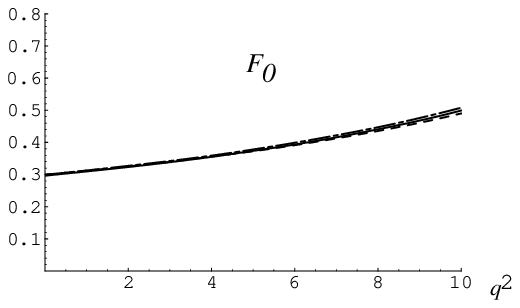}}
\caption{The $B\to \pi$ form factors $F^{B\pi}_+$ and $F^{B\pi}_0$
as functions of $q^2$ (GeV$^2$). The results by Gaussian,
exponential and KKQT type B distribution amplitude are 
shown in solid, dot and dot-dashed line, respectively.
} \label{Bpi-val}
\end{figure}

We present the numerical results of the $B\rightarrow \pi$ transition
form factors given above.
The default values for inputs are given as follows:
\begin{equation}
\begin{array}{ccc}
f_B = 190 \text{ MeV}\ , & f_\pi = 130 \text{ MeV}\ , &
\Lambda_{\rm QCD} = 250 \text{ MeV ($N_f=4$)}\ ,\\
c = 0.3\ , & m_0 = 1.4 \text{ GeV}\ , &\\
a_2 = 0.44 \ , & a_4 = 0.25 \ , & a_{2p}= 0.43 \ ,\\
 a_{4p}=0.09 \ , & a_{2t}= 0.55/6\ . &
\end{array}
\label{Bpi-inputs}
\end{equation}
The parameters in $B$ meson distribution amplitudes are chosen as 
$\omega_{KLS} = 0.38$, $\omega_{GN}=0.36$ and $\Lambda_{KKQT}/M_B = 0.094$
so that we have $F^{B\pi}_{+,0}(0)\cong 0.3$, which is reasonable 
in comparison with the sum-rule results\cite{PB4}.
We neglect the scale dependence of parameters
$m_0$ and $a_2\;, \dots , a_{2t}$ in the default calculation.
Its effect shall be discussed later.
Monte-Carlo method is used to evaluate the numerical integrals. We have set
the number of samples so that the statistical errors in
Monte-Carlo integrations may be less than 0.1\%.
The values of two form factors should be equal at $q^2=0$. 
The PQCD results becomes unreliable gradually at slow recoil.
Our results of $F^{B\pi}_+(q^2)$ and $F^{B\pi}_0(q^2)$ for $q^2 = 0
\sim 10$ GeV$^2$ are shown in Table~\ref{tbl-def} and Fig.~\ref{Bpi-val}.
It can be seen that the $q^2$ dependences are almost same irrespective of
the choice of the $B$ meson distribution amplitude. The difference is at most
4 \% at $q^2 =10$ GeV${}^2$.
The ratio of each contribution  from $\phi_\pi$, $\phi^p_\pi$ and
$\phi^t_\pi$ to the total value of $F^{B\pi}_+(0)$ is given in
Table~\ref{tbl-Bpicont}. It shows that the twist-3 contribution
is important as explained in \cite{TLS}.
\begin{table}[ht]
Gaussian type with $\omega_{KLS}=0.38$:
\\
\begin{small}
\begin{tabular}{c|ccccccccccc}\hline
$q^2$ (GeV${}^2$)&0.0&1.0&2.0&3.0&4.0&5.0&6.0&7.0&8.0&9.0&10.0\\\hline
$F^{B\pi}_0$ &0.297&0.310&0.324&0.339&0.355&0.374&0.393&0.416&0.441&0.468&0.499\\
$F^{B\pi}_+$ &0.297&0.321&0.347&0.377&0.411&0.450&0.494&0.546&0.605&0.674&0.756\\\hline
\end{tabular}
\end{small}

\vskip3mm
Exponential type with $\omega_{GN}=0.36$:
\\
\begin{small}
\begin{tabular}{c|ccccccccccc}\hline
$q^2$ (GeV${}^2$)&0.0&1.0&2.0&3.0&4.0&5.0&6.0&7.0&8.0&9.0&10.0\\\hline
$F^{B\pi}_0$ &0.300&0.312&0.325&0.339&0.356&0.373&0.391&0.413&0.436&0.461&0.490\\
$F^{B\pi}_+$ &0.300&0.323&0.349&0.378&0.412&0.449&0.492&0.542&0.599&0.665&0.743\\\hline
\end{tabular}
\end{small}

\vskip3mm
KKQT type with {$\displaystyle \frac{\Lambda_{KKQT}}{m_B}=0.094$}:
\\
\begin{small}
\begin{tabular}{c|ccccccccccc}\hline
$q^2$ (GeV${}^2$)&0.0&1.0&2.0&3.0&4.0&5.0&6.0&7.0&8.0&9.0&10.0\\\hline
$F^{B\pi}_0$ &0.299&0.313&0.327&0.342&0.359&0.378&0.399&0.422&0.447&0.476&0.508\\
$F^{B\pi}_+$ &0.299&0.324&0.351&0.381&0.415&0.456&0.501&0.554&0.615&0.686&0.770\\\hline
\end{tabular}
\end{small}
\caption{Numerical outputs of $F^{B\pi}_0(q^2)$ and $F^{B\pi}_+(q^2)$}
\label{tbl-def}
\end{table}
\begin{table}[h]
\begin{small}
\begin{tabular}{c|ccc}\hline
&Gaussian&Exponential&KKQT\\\hline
$\phi_\pi$ (\%)  &40&37&40\\
$\phi^p_\pi$ (\%)&47&51&46\\
$\phi^t_\pi$ (\%)&13&12&14\\\hline
\end{tabular}
\end{small}
\caption{The contributions from $\phi_\pi$, $\phi^p_\pi$ and $\phi^t_\pi$
to the total value of $F^{B\pi}(0)$.}
\label{tbl-Bpicont}
\end{table}

\subsection{Parameters in distribution amplitudes}
\begin{figure}
\resizebox{7cm}{!}{
\includegraphics{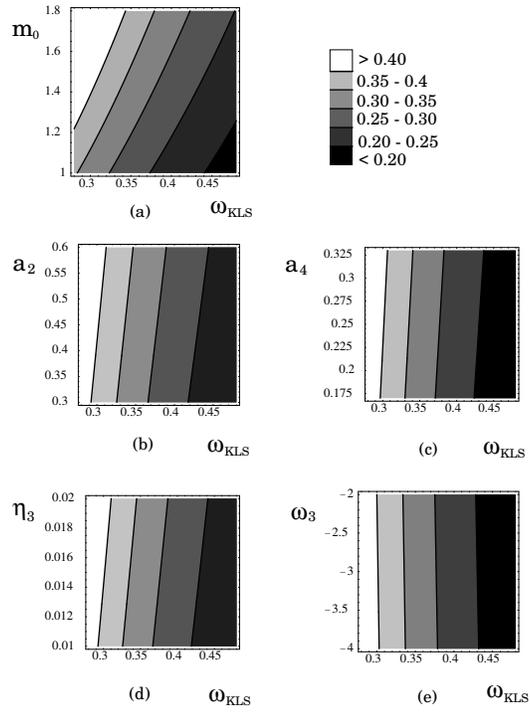}
}
\caption{
Contour plots of $F^{B\pi}(0)$.
(a) $\omega_{KLS}$ - $m_0$,
(b) $\omega_{KLS}$ - $a_2$ (c) $\omega_{KLS}$ - $a_4$,
(d) $\omega_{KLS}$ - $\eta_3$ (e) $\omega_{KLS}$ - $\omega_3$.
The values of $F^{B\pi}(0)$ are shown by using shades as given in the sample .
}
\label{w-mfig-our}
\end{figure}

Each of the $B$ meson distribution amplitudes, 
Eqs.~(\ref{ourBwv}), (\ref{GNwv}) and (\ref{KKQTwv}), adopted
in the previous calculation has only one parameter $\omega_{KLS}$,
$\omega_{GN}$ and $\Lambda_{KKQT}$, respectively. The pion
distribution amplitudes, Eqs.~(\ref{pioa})-(\ref{pioc}), contain 5
parameters, $m_0^\pi$, $a_1$, $a_2$, $\eta_3$ and $\omega_3$.
We study how the numerical outputs of the form factors at $q^2=0$
vary with these parameters. The form factor at $q^2=0$ can be
rewritten by factoring out the parameters in the pion distribution
amplitudes as
\begin{eqnarray}
F^{B\pi}(0) &=& F^{A0}(X) + a_2 F^{A2}(X)
+ a_4 F^{A4}(X) 
+ \frac{m_0}{m_B}\left[ F^{P0}(X) + a_{2p}F^{P2}(X) \right.\\
&&
\left. 
+ a_{4p} F^{P4}(X) + F^{T0}(X) +
a_{2t} F^{T2}(X) \right],
\end{eqnarray}
where $X=\omega_{KLS}$, $\omega_{GN}$ or $\Lambda_{KKQT}$.
The functions $F^{A0}$, $F^{A2}$ \dots \ $F^{T2}$ do not depend on
the pion parameters.
\begin{description}
\item[Gaussian type:]
In the case of Gaussian type $B$ meson distribution amplitude,  
the $\omega_{KLS}$ dependence can be well
approximated within 1\% precision for $0.28 \le \omega_{KLS} \le 0.48$
by the following formulae;
\begin{eqnarray}
F^{A0}(\omega_{KLS})&=& 0.0623 - 0.175\,(\omega_{KLS} -0.38 )  + 0.382\,(\omega_{KLS} -0.38 )^2
        - 0.784\,(\omega_{KLS} -0.38 )^3\;, \nonumber\\
F^{A2}(\omega_{KLS})&=& 0.0860 - 0.246\,(\omega_{KLS} -0.38 )  + 0.455\,(\omega_{KLS} -0.38 )^2
         - 0.634\,(\omega_{KLS} -0.38 )^3\;, \nonumber\\
F^{A4}(\omega_{KLS})&=& 0.0784 - 0.231\,(\omega_{KLS} -0.38 )  + 0.417\,(\omega_{KLS} -0.38 )^2
         - 0.242\,(\omega_{KLS} -0.38 )^3\;, \nonumber\\
F^{P0}(\omega_{KLS})&=& 0.446 - 1.98\,(\omega_{KLS} -0.38 )  +  6.47\,(\omega_{KLS} -0.38 )^2
   - 17.5\,(\omega_{KLS} -0.38 )^3\;,  \nonumber\\
F^{P2}(\omega_{KLS})&=& 0.153 - 0.563\,(\omega_{KLS} -0.38 )  + 1.34,(\omega_{KLS} -0.38 )^2
   - 2.05\,(\omega_{KLS} -0.38 )^3\;, \nonumber\\
F^{P4}(\omega_{KLS})&=& 0.0825 - 0.288\,(\omega_{KLS} -0.38 )  + 0.591\,(\omega_{KLS} -0.38 )^2
   - 0.471\,(\omega_{KLS} -0.38 )^3\;, \nonumber\\
F^{T0}(\omega_{KLS})&=& 0.109 - 0.332\,(\omega_{KLS} -0.38 )  + 0.635\,(\omega_{KLS} -0.38 )^2
   - 0.734\,(\omega_{KLS} -0.38 )^3\;, \nonumber\\
F^{T2}(\omega_{KLS})&=& 0.441 - 1.37\,(\omega_{KLS} -0.38 )  + 2.64\,(\omega_{KLS} -0.38 )^2
   - 3.86\,(\omega_{KLS} -0.38 )^3\; .
\end{eqnarray}
The chiral mass $m_0 = m_\pi^2/(m_u + m_d)$ plays an important
role in hadron dynamics. It gives penguin enhancement in $B$ meson
non-leptonic decays as pointed out in \cite{KLS}. It is
essential for the form factor calculation to take into account of the
important higher-twist contributions\cite{TLS}. The  chiral mass
$m_0$ enters in Eqs.~(\ref{FormBpi2}) and (\ref{FormBpi3}) as
$r_\pi = m_0/m_B$ and in the parameter $\rho_\pi$ of Gegenbauer
polynomials in the pion distribution amplitudes. 
(See Eqs.~(\ref{gegenap})-(\ref{defrho}).) 
The $r_\pi$ dependence of the from factor is
linear, while the parameter $\rho_\pi$ in pion distribution
amplitudes depends linearly on $1/m_0$. The $m_0$ dependence of
$a_{2p}$, $a_{4p}$ and $a_{2t}$ through $\rho^2$ can be neglected
since $\rho_\pi^2 = O(10^{-2})$.
The $\omega_{KLS}$ - $m_0$ dependence of $F^{B\pi}(0)$ is shown in
Fig. \ref{w-mfig-our} (a), where other parameters are fixed to the
default values. The dependence on other inputs are also shown in
Figs. \ref{w-mfig-our} (b)-(e). 
\item[Exponential type:]
The similar calculation is done in the case of the exponential
type $B$ meson distribution amplitude. 
The $\omega_{GN}$ dependence is obtained for $0.26 \le \omega_{GN} \le 0.46$. 
The approximation formulae of $F^{A0} \sim F^{T2}$
are given in the appendix A.
The result is shown in Fig. \ref{w-mfig-GN}.
\item[KKQT type:]
The result in the case of the KKQT type $B$ meson distribution amplitude
is shown in Fig. \ref{w-mfig-KKQT}.
The $\Lambda_{KKQT}$ dependence is obtained 
for $0.074 \le \Lambda_{KKQT}/M_B \le 0.114$. 
The approximation formulae of $F^{A0} \sim F^{T2}$
are given in the appendix A.
\end{description} 
\begin{figure}
\resizebox{7cm}{!}{
\includegraphics{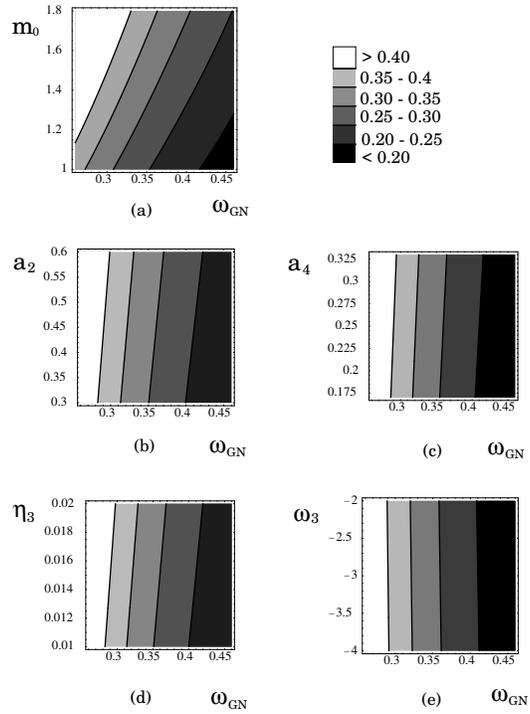}
}
\caption{
Contour plots of $F^{B\pi}(0)$.
(a) $\omega_{GN}$ - $m_0$,
(b) $\omega_{GN}$ - $a_2$ (c) $\omega_{GN}$ - $a_4$,
(d) $\omega_{GN}$ - $\eta_3$ (e) $\omega_{GN}$ - $\omega_3$.
The values of $F^{B\pi}(0)$ are shown by using shades as given in the sample .
}
\label{w-mfig-GN}
\end{figure}
\begin{figure}[h]
\resizebox{7cm}{!}{
\includegraphics{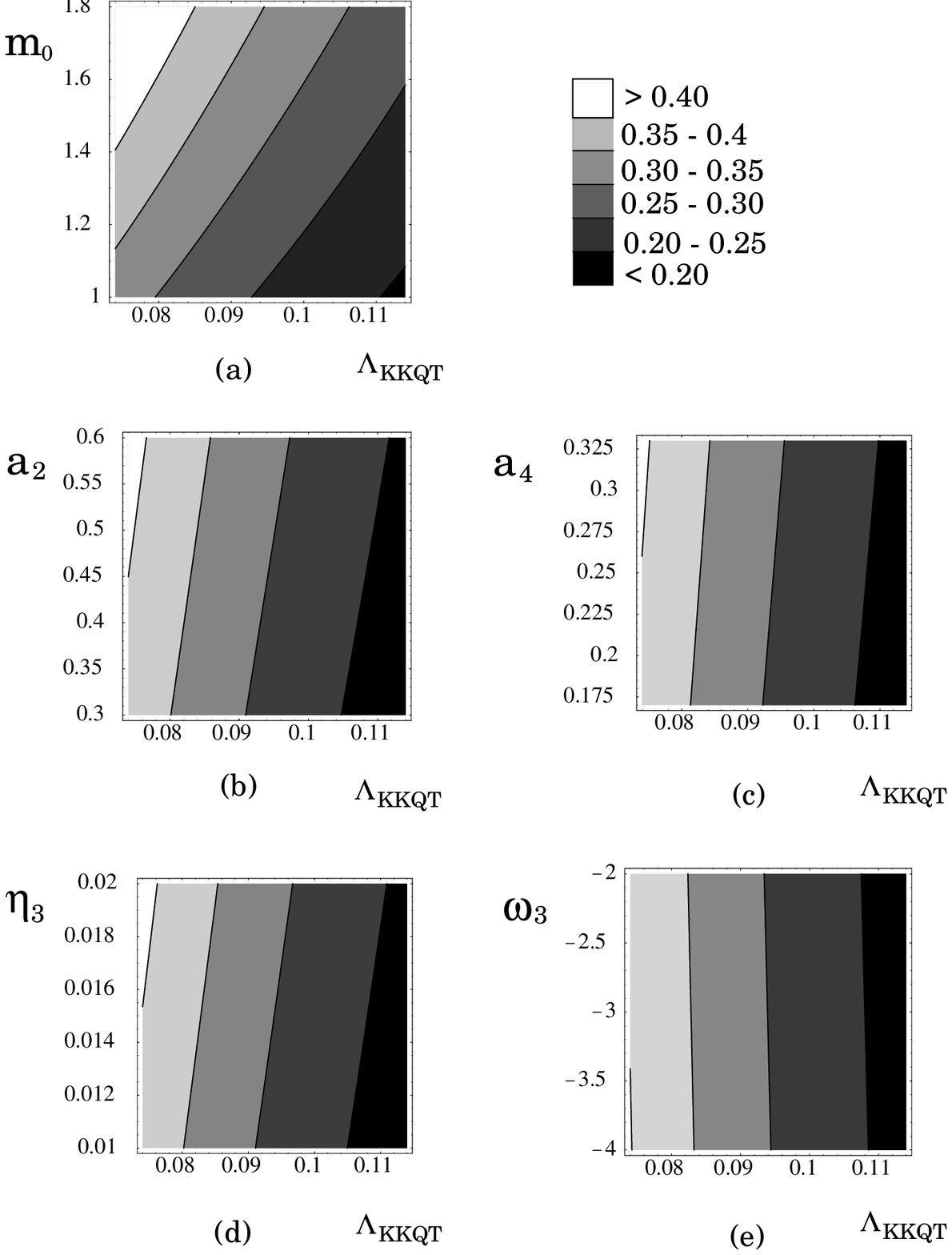}
}
\caption{
Contour plots of $F^{B\pi}(0)$.
(a) $\Lambda_{KKQT}$ - $m_0$,
(b) $\Lambda_{KKQT}$ - $a_2$ (c) $\Lambda_{KKQT}$ - $a_4$,
(d) $\Lambda_{KKQT}$ - $\eta_3$ (e) $\Lambda_{KKQT}$ - $\omega_3$.
$\Lambda_{KKQT}$ is given in unit of $M_B$.
The values of $F^{B\pi}(0)$ are shown by using shades as given in the sample .
}
\label{w-mfig-KKQT}
\end{figure}

These figures show that $F^{B\pi}(0)$
depends most significantly on $m_0$ and a $B$ meson distribution 
amplitude parameter ($\omega_B$, $\omega_{GN}$ or  $\Lambda_{KKQT}$). 
The change of
other parameters within reasonable range affects on $F^{B\pi}(0)$
at most 10\%.

The $B$ meson decay constant, $f_B$, is concerned solely with the
normalization of $B$ meson wave function in the form factor
calculation. The normalization constant $N_B$ enters linearly in
our calculation.  So if $f_B$ changes to $f_B +\Delta f_B$, the
output changes to $(1 + \Delta f_B/f_B)$ times the original value.
This is also the case in the calculation of non-leptonic decays
\subsection{Intrinsic $b$ dependence}
We investigate the uncertainty from the intrinsic $b$
dependence of light meson distribution amplitudes, which are
advocated by Kroll et al.\cite{bdepK}. The $b$ dependence of 
pion is taken to be the following form \cite{bdepK},
\begin{equation}
\exp\left[- \frac{x(1-x)b^2}{4a_\pi^2} 
\right]\;,
\end{equation} 
where $a_\pi$ is the transverse size parameter of the pion.
We take $a_\pi^{-1} \simeq \sqrt{8}\pi f_\pi$ here.  
The variation of $F^{B\pi}(0)$ under the influence of 
the above $b$ dependence of pion distribution amplitude is 
shown in Table~\ref{tbl-bdep}. The effects of intrinsic $b$ dependence is 
estimated to be about 10\%
or less.
\begin{table}[h]
\begin{tabular}{r|ccc}\hline
$a_\pi^{-1}$ & Gaussian & Exponential & KKQT \\\hline
without $b$ dependence  & 0.297 & 0.300 & 0.299 \\
$0.8 \times \sqrt{8}\pi f_\pi$&0.284& 0.285 & 0.286 \\
$\sqrt{8}\pi f_\pi$& 0.277&0.278&0.279 \\
$1.2 \times \sqrt{8}\pi f_\pi$& 0.269&0.269&0.272 \\\hline
\end{tabular}
\caption{$b$ dependence of $F^{B\pi}(0)$}
\label{tbl-bdep}
\end{table} 
\subsection{Evolution effect}
\subsubsection{Gegenbauer coefficients}
The Gegenbauer coefficients in the light meson distribution amplitudes
depend on the energy scale. In the PQCD calculation it
evolves with the scale $1/b$ governed 
by $[\alpha_s(1/b)/\alpha_s(\mu_0)]^{\gamma/b}$ ($b=11-2N_f/3$), where
$\mu_0$ represents the initial scale the evolution starts
with, and $\gamma$ is an anomalous dimension \cite{BB}. We have
investigated this evolution effect. Calculations are made by taking 
the evolution effect into account.
It can be seen from Table~\ref{tbl-evol} that the effect is 
about 10\%, and can be covered by the theoretical uncertainty 
from the variation of the Gegenbauer coefficients.
\begin{table}
\begin{tabular}{r|ccc}\hline
$\mu_0$ (GeV)& Gaussian & Exponential & KKQT \\\hline
no evolution & 0.297 & 0.300 & 0.299 \\
0.5 &0.347&0.345&0.352\\
1.0 &0.294&0.297&0.298\\
1.5 &0.278&0.282&0.280\\\hline
\end{tabular}
\caption{The evolution effect on $F^{B\pi}(0)$}
\label{tbl-evol}
\end{table}
\subsubsection{$\Lambda_{\rm QCD}$}
The QCD coupling constant $\alpha_S$ appears explicitly and implicitly
through the resummation factor $S$ in Eq.~(\ref{evol}).
The QCD scale $\Lambda_{\rm QCD}$ determines $\alpha_S$. Let us
see how the form factor values varies depending on $\Lambda_{\rm QCD}$.
The result is given in Table~\ref{tbl-QCD}, which shows that change in
the form factor values is about 3\%  for
$200 \text{ MeV} \le \Lambda_{\rm QCD}  \le 300 \text{ MeV}$.
\begin{table}[h]
\begin{tabular}{r|ccc}\hline
$\Lambda_{\rm QCD}$ (MeV)& Gaussian & Exponential & KKQT \\\hline
200& 0.299&0.308&0.310\\
225& 0.298&0.305&0.306\\
250& 0.297&0.300&0.299\\
275& 0.293&0.294&0.294\\
300& 0.289&0.288&0.286\\\hline
\end{tabular}
\caption{$\Lambda_{\rm QCD}$ dependence of $F^{B\pi}(0)$}
\label{tbl-QCD}
\end{table}
\subsubsection{Hard scales}
The scale of $\alpha_s$ in the expression of the form factors is
determined in Eq.~(\ref{scale_t}). This choice is not unique,
because the next leading order correction has not been calculated. 
There is another candidate of the scale:
\begin{equation}
t^{(1)} = t^{(2)}= {\rm
max}(\sqrt{x_1\eta}m_B,\sqrt{x_2\eta}m_B,1/b_1,1/b_2)\;,
\label{another_t}
\end{equation}
The change of the value of $F^{B\pi}(0)$ under the choice of 
hard scale is shown in Table~\ref{tbl-t1t2}. 
For reference, we also show the value in the case of the 
fixed hard scales; $t^{(1)} = t^{(2)}= M_B/2$, $M_B$, $2M_B$.
The result shows that $ F^{B\pi}(0)$ changes about 10\% or less depending on the
choice of the form of the scale $t^{(1,2)}$.
\begin{table}[h]
\begin{tabular}{l|ccc}\hline
& Gaussian & Exponential & KKQT \\\hline
original & 0.297&0.300&0.299\\
Eq.~(\ref{another_t}) & 0.288&0.290&0.289\\
fixed $t^{(1)} = t^{(2)}=M_B/2$& 0.286 & 0.288 &0.288\\
fixed $t^{(1)} = t^{(2)}=M_B$& 0.276 & 0.277 & 0.277\\
fixed $t^{(1)} = t^{(2)}=2M_B$& 0.269 & 0.270 & 0.270\\
\hline
\end{tabular}
\caption{Scale choice dependence of $ F^{B\pi}(0)$}
\label{tbl-t1t2}
\end{table}
\subsubsection{Threshold resummation factors}
There is a source of
theoretical uncertainty from the threshold resummation factor $c$ 
in Eq.~(\ref{trs}).
Note that this uncertainty, whose property differs from others
like $m_0$, is not due to an unknown parameter, but to our
parameterization. In principle, we can adopt the exact resummation
result, such that no theoretical uncertainty is associated with it.
\subsection{Sub-leading contribution}
\subsubsection{$O(\Lambda_{hadron}/m_B)$ terms}
The formulae of the form factors $f_1$ and $f_2$ are the leading order
results where the terms proportional to $x_1 \sim \Lambda_{hadron}/m_B$ are
neglected. If we do not neglect them, the following terms are added.
\begin{eqnarray}
\Delta f_1&=&16\pi m_B^2C_F\int dx_1dx_2\int b_1db_1
b_2db_2\phi_B(x_1,b_1) x_1 (\eta \phi_\pi - 2 r_\pi \phi^p_\pi)
\nonumber\\
&&\times
 E(t^{(2)})h(x_2,x_1,b_2,b_1)\;,
\label{x1Bpi1}\\
\Delta f_2&=&16\pi m_B^2C_F\int dx_1dx_2\int b_1db_1
b_2db_2\phi_B(x_1,b_1) x_1\left( -\phi_\pi + \frac{2
r_\pi}{\eta}\phi^p_\pi \right)
\nonumber\\
&&\times
 E(t^{(2})h(x_2,x_1,b_2,b_1)\;,
\label{x1Bpi2}
\end{eqnarray}
The numerical outputs of the above quantities are given 
in Table~\ref{tbl-deltaf}. 
It can be fond that the sub-leading
contribution from $x_1$ terms is about 4\% of the leading value.
\begin{table}[h]
\begin{tabular}{ccc}\hline
 Gaussian & Exponential & KKQT \\\hline
 0.040 & 0.035& 0.041 \\\hline
\end{tabular}
\caption{$\Delta F^{B\pi}(0)/F^{B\pi}(0)$} 
\label{tbl-deltaf}
\end{table}

The leading order results of PQCD calculation are obtained under the 
approximation of $M_B = m_b$. If we do not take this approximation 
the formulae become as follows;
\begin{eqnarray}
f_1&=&16\pi m_B^2C_Fr_\pi\int dx_1dx_2\int b_1db_1
b_2db_2\phi_B(x_1,b_1) [(1+r_B)\phi_\pi^p(x_2)-(1-r_B)\phi_\pi^t(x_2)]
\nonumber\\
&&\times
 E(t^{(1)})h_B(x_1,x_2,b_1,b_2)\;,
\label{FormBpi2b}\\
f_2&=&16\pi m_B^2C_F\int dx_1dx_2\int b_1db_1
b_2db_2\phi_B(x_1,b_1)
\nonumber\\
&\times& \Biggl\{\left[\phi_\pi(x_2)(1+x_2\eta -r_B)
+2r_\pi\left((1-\frac{r_B}{2})(\frac{1}{\eta} -x_2 )\phi_\pi^t(x_2) -x_2\phi_\pi^p(x_2)
 \right)\right]E(t^{(1)})h_B(x_1,x_2,b_1,b_2)
\nonumber\\
& &+ 2r_\pi\phi_\pi^p E(t^{(2})h(x_2,x_1,b_2,b_1)\Biggr\}\;,
\label{FormBpi3b}
\end{eqnarray}
where $r_B = \bar\Lambda/M_B = (M_B -m_b)/M_B$.
The hard function $h_B$ is given as
\begin{eqnarray}
h_B(x_1,x_2,b_1,b_2)&=&S_t(x_2)K_{0}\left(\sqrt{x_1x_2\eta}m_Bb_1\right)
\nonumber \\
& &\times \theta (x_2\eta - 2r_B)\left[\theta(b_1-b_2)
K_0\left(\sqrt{x_2\eta -2r_B}m_B
b_1\right)I_0\left(\sqrt{x_2\eta -2r_B}m_Bb_2\right)\right.
\nonumber \\
& &\left.+\theta(b_2-b_1)K_0\left(\sqrt{x_2\eta -2r_B}m_Bb_2\right)
I_0\left(\sqrt{x_2\eta -2r_B}m_Bb_1\right)\right]\;. \label{dh2}
\end{eqnarray}
The outputs of $F^{B\pi}(0)$ with the above formulae are given in Table~\ref{tbl-Bb}.
Comparing this result with the leading order one, we find that the effect of 
this approximation is about 2\%.
\begin{table}[h]
\begin{tabular}{ccc}\hline
 Gaussian & Exponential & KKQT \\\hline
 0.302 & 0.305& 0.304 \\\hline
\end{tabular}
\caption{$ F^{B\pi}(0)$ without taking $M_B=m_b$}
\label{tbl-Bb}
\end{table}

\subsubsection{Another component of $B$ distribution amplitudes}
\label{anoB}
The $B$ meson distribution amplitude in fact consists of
two components\cite{GN}; 
\begin{equation}
\Phi_B =
-\frac{i}{\sqrt{2N_c}}(\not P+m_B)\gamma_5 \left[\not{v_+}
\phi_B^{+}(k) +\not{v_-}\phi_B^{-}(k)\right] \;,
\end{equation}
where $v = P/m_B = v_+ + v_-$ with $v_+=((v^0+v^3)/\sqrt{2},0,0_T)$ and 
$v_-=(0,(v^0-v^3)/\sqrt{2},0_T)$. The spatial direction of the 
velocity $v$ is taken along the third direction ($v^1 = v^2 =0$). 
By using the identity
$(\not P+m_B)\gamma_5 (1+\not v) =0$, we can add an arbitrary
function $f$ in the above expression;
\begin{eqnarray}
\Phi_B 
&\propto&(\not P+m_B)\gamma_5 \left[\not{v_+}\phi_B^{+} +
\not{v_-}\phi_B^{-}\right]
\nonumber\\
&=& (\not P+m_B)\gamma_5 \left[\not{v_+}\phi_B^{+} 
+\not{v_-}\phi_B^{-} +
(1+\not v) f
\right]\nonumber\\
&=& (\not P+m_B)\gamma_5 \left[f +\not{v_+}(\phi_B^{+} + f) +
\not{v_-}(\phi_B^{-} +f)
\right]\nonumber\\
&=& -(\not P+m_B)\gamma_5 \left[ (f + \phi_B^{+} + \phi_B^{-})
+\not{v_-}(\phi_B^{+} + f) +\not{v_+}(\phi_B^{-} +f)
 \right] \ .  \label{bwv-add}
\end{eqnarray}
In the rest frame of $B$ meson, $v^0=1$ and other components of 
$v$ vanish, so that we have
\begin{equation}
\Phi_B = \frac{i}{\sqrt{2N_c}}(\not P+m_B)\gamma_5 \left[(f + \phi_B^{+} + \phi_B^{-})
+ \frac{\not n}{\sqrt{2}} (\phi_B^{+} + f) +
\frac{\not {\bar n}}{\sqrt{2}}(\phi_B^{-} +f)
\right] \ , 
 \label{bwv-add-rest}
\end{equation}
where $n = (0,1,0_T)$ and $\bar n = (1,0,0_T)$. 
We have so far considered the contribution from the first term 
alone by choosing $f = -\phi_B^{+}$ or $f=-\phi_B^{-}$, and 
that from the rest of the terms has been neglected.
Here we estimate the contribution from the rest of the 
$B$ meson distribution amplitude components.

A care is necessary in choosing $\phi_B^\pm$ in the rest 
frame of $B$ meson where we need to distinguish ``+" direction. 
In \cite{GN} and \cite{KKQT}, the coordinate of the light 
quark in $B$ meson is denoted as $z$ which is on the light-cone,
$z^2=z^+z^-=0$. 
\begin{equation}
\langle 0|
\bar q (z)\Gamma h_v(0)|\bar B(p)\rangle 
= -\frac{if_B M_B}{2} \text{Tr}\left[
\gamma_5 \Gamma \frac{1+\not v}{2}\left\{ 
\tilde \phi_+(v\cdot z) - 
\not z \frac{\tilde\phi_+(v\cdot z) - \tilde\phi_-(v\cdot z)}{2v\cdot z} 
\right\}\right], 
\end{equation} 
where $\tilde\phi_\pm$ are the Fourier transforms of $\phi_\pm$. 
The function $\phi_B^+$ is defined as the distribution
amplitude associated with $v^+$, so that it becomes
the leading distribution amplitude in the 
limit $t=v\cdot z =v^+z^- + v^-z^+\to \infty$. 
Since $z^+z^-=0$, $z^+= 0$ and $z^- \neq 0$ are taken in their treatment.
Then the momentum of the light quark, $k$,  should be 
taken along ``+" direction, so that $z\cdot k \neq 0$.
We have taken the light quark momentum along ``--" direction 
in the calculation of  Eqs.~(\ref{FormBpi2}) and (\ref{FormBpi3})  
\cite{TLS}. Then $\phi^-$ is the leading distribution amplitude, and 
our choice corresponds to $f = -\phi_B^{+}$.

Let us express $\phi_B = f + \phi_B^{+} + \phi_B^{-}$, 
$\phi_B^n = \phi_B^{+} + f$ and  $\phi_B^{\bar n} = \phi_B^{-} + f$.
The contribution from $\phi_B$ is given in Eqs.~(\ref{FormBpi2}) and 
(\ref{FormBpi3}). The contribution from $\phi_B^n$ is given as
\begin{eqnarray}
f_1^n&=& 0 \ ,\label{FormBpi1n}\\
f_2^n&=&-16\pi m_B^2C_F\int dx_1dx_2\int b_1db_1
b_2db_2\phi_B^n(x_1,b_1)
\nonumber\\
&\times& \Biggl\{\left[ x_2\eta \phi_\pi(x_2)
+r_\pi(\frac{1}{\eta} -x_2 )(\phi_\pi^p(x_2) +\phi_\pi^t(x_2) )
\right]E(t^{(1)})h(x_1,x_2,b_1,b_2)
\nonumber\\
& &+ 2r_\pi\phi_\pi^p E(t^{(2})h(x_2,x_1,b_2,b_1)\Biggr\}\;.
\label{FormBpi2n}
\end{eqnarray}
The contributions from $\phi_B^{\bar n}$ 
is given as
\begin{eqnarray}
f_1^{\bar n}&=&-16\pi m_B^2C_Fr_\pi\int dx_1dx_2\int b_1db_1
b_2db_2\phi_B^{\bar n}(x_1,b_1) [\phi_\pi^p(x_2)-\phi_\pi^t(x_2)]
\nonumber\\
&&\times  
 E(t^{(1)})h(x_1,x_2,b_1,b_2)\;,
\label{FormBpi1nbar}\\
f_2^{\bar n}&=&-16\pi m_B^2C_F\int dx_1dx_2\int b_1db_1
b_2db_2\phi_B^{\bar n}(x_1,b_1) 
\nonumber\\
&\times& 
\left[\phi_\pi(x_2)
-r_\pi(\frac{1}{\eta} +x_2 )\phi_\pi^p(x_2) 
+ r_\pi(\frac{1}{\eta} -x_2 ) \phi_\pi^t(x_2)
\right]E(t^{(1)})h(x_1,x_2,b_1,b_2)
\;.
\label{FormBpi2nbar}
\end{eqnarray}
Note that the sum of contributions
from $\phi_B$, $\phi^{n}_B$ and $\phi^{\bar n}_B$
vanishes if $\phi_B = \phi^{n}_B = \phi^{\bar n}_B$.
It is because
\begin{eqnarray}
[\not P + m_B ]\gamma_5 \left[
\phi_B(x) + \frac{\not n}{\sqrt{2}}\phi_B^{n}(x)
+ \frac{\not {\bar n}}{\sqrt{2}}\phi_B^{\bar n}(x)
\right] &=&
[\not P + m_B ]\gamma_5 [
1 + \frac{\not n}{\sqrt{2}} + \frac{\not {\bar n}}{\sqrt{2}}]\phi_B(x)
\nonumber \\
&=& m_B[1 + \not v] \gamma_5 [1 + \not v]\phi_B(x) =0 \;.
\end{eqnarray}

We need $\phi_B^{+}$ and $\phi_B^{-}$ to calculate the numerical 
values of these contributions. The candidates of the leading 
distribution amplitude, $\phi_B^{-}$ in this case,  are 
already given in Eqs.(\ref{ourBwv}) - (\ref{KKQTwv}). 
For KKQT type distribution amplitude, $\phi_B^{+}$ is derived in 
\cite{KKQT}. 
(Note that the $\phi_B^{-}$ in \cite{KKQT} corresponds to $\phi_B^{+}$ here.)
\begin{equation}
 \phi_B^{+(KKQT)}(x,b) = 
N_B\; (2\frac{\Lambda_{KKQT}}{m_B} - x )
\theta(x)\theta(\frac{2\Lambda_{KKQT}}{m_B} -x)
J_0 \left(
        b\sqrt{x(\frac{2\Lambda_{KKQT}}{m_B} -x)}\right)
\ .\label{KKQT-mns}
\end{equation} 
The $x$ dependence of the candidate in the case of exponential type 
is proposed in \cite{GN}. We add the same $b$ dependence as in Eq.(\ref{GNwv}).
\begin{equation}
\phi_B^{+(GN)}(x,b)=N_B^{GN}\left(\frac{\omega_{GN}}{m_B} \right)
  \exp\left[-\frac{xm_B}{\omega_{GN}}
\right]\frac{1}{1 + (b\omega_{GN})^2} \;. \label{GNwv-mns}
\end{equation} 
As for the Gaussian type case, a candidate of $\phi_B^{+}$ is  
proposed in \cite{WY} by solving the equation of motions given in 
\cite{KKQT} with $(\phi_B^{+} + \phi_B^{-})/2= \phi_B^{KLS}(x,b)$. 
Here we take $\phi_B^{-} = \phi_B^{KLS}(x,b)$, and put it into 
the equation of motions. The details are given in the Appendix B. 
The result is as follows; 
\begin{eqnarray}
\phi_B^{+(KLS)}(x,b)&=& N_B^{KLS}\;
\frac{\omega_{KLS}^2}{m_B^4} \left[
\exp\left[-\frac{1}{2}\left( \frac{x m_B}{\omega_{KLS}} \right)^2\right]
\left\{m_B^2 (1-x)^2 + 2 \omega_{KLS}^2 \right\}\right. \nonumber\\
&& \left.
 + \sqrt{2\pi}m_B \text{Erf}\left(\frac{x m_B }{\sqrt{2}\omega_{KLS}} \right)
\right] 
\exp\left[-\frac{1}{2}( \omega_{KLS} b )^2\right] + C \;, \label{gaussmns}
\end{eqnarray} 
where the constant $C$ is chosen so that $\phi_B^{+(KLS)}(1,b)=0$.

\begin{figure}[h]
\resizebox{5cm}{!}{\includegraphics{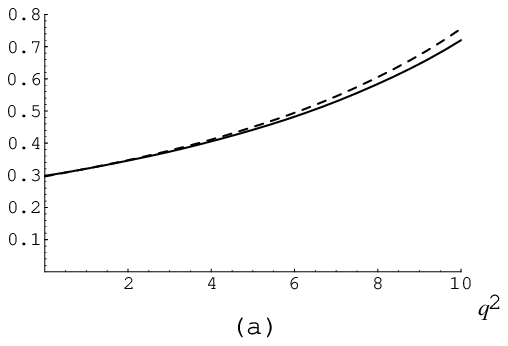}}

\resizebox{5cm}{!}{\includegraphics{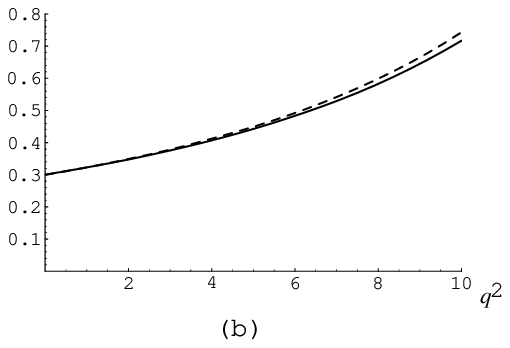}}

\resizebox{5cm}{!}{\includegraphics{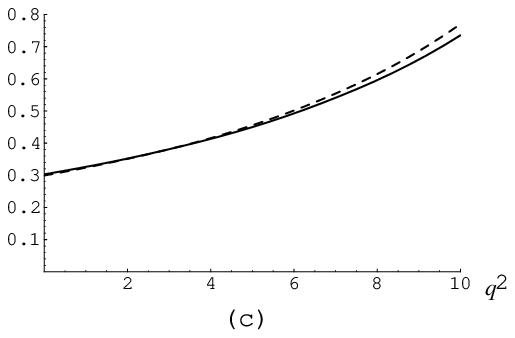}}
\caption{The $B\to \pi$ form factors $F^{B\pi}_+$ as functions of $q^2$ 
(GeV$^2$) for Gaussian type (a), exponential type (b) and 
KKQT type (c) distribution amplitudes. 
The results with the leading contribution only 
with a suitable choice of the parameter
are shown in dashed lines, while
those with both contributions are shown in solid lines. 
} \label{Bpi-sub}
\end{figure}

%
We have investigated the contributions from $\phi_B^{\bar n}$ components 
under the choice of $f=-\phi_B^+$. 
The results of $F^{B\pi}(0)$  with both $\phi_B^+$ and $\phi_B^-$ contributions
are shown in Table~\ref{tbl-kkqt0}. 
It can be found that $\omega_{KLS} = 0.45$, $\omega_{GN} = 0.42$ or  
$\Lambda_{KKQT}/m_B = 0.12$ gives $F^{B\pi}(0)\cong 0.3$.
\begin{table}[h]
\begin{tabular}{c|ccccc}\hline
$\omega_{KLS}$  & 0.43 & 0.44 & 0.45 & 0.46 & 0.47 \\\hline
$F^{B\pi}(0)$ & 0.317 & 0.308 & 0.298 & 0.289 & 0.281 \\\hline
\end{tabular}

\vskip3mm
\begin{tabular}{c|ccccc} \hline
$\omega_{GN}$  & 0.40 & 0.41 & 0.42 & 0.43 & 0.44 \\\hline
$F^{B\pi}(0)$ & 0.321 & 0.310 & 0.300 & 0.291 & 0.282 \\\hline
\end{tabular}

\vskip3mm
\begin{tabular}{c|ccccc} \hline
$\Lambda_{KKQT}/m_B$  & 0.10 & 0.11 & 0.12 & 0.13 & 0.14 \\\hline
$F^{B\pi}(0)$ & 0.374 & 0.336 & 0.303 & 0.274 & 0.249  \\\hline
\end{tabular}
\caption{The value of $F^{B\pi}(0)$ for 
$\omega_{KLS}$, $\omega_{GN}$ and 
$\Lambda/m_B$ by using
Gaussian, exponential and KKQT type distribution amplitudes, respectively.}
\label{tbl-kkqt0}
\end{table}
The $q^2$ dependence of $F^{B\pi}_+$ with both contributions is
shown in Fig.\ref{Bpi-sub}. The results with the leading contribution only
($\omega_{KLS} = 0.38$, $\omega_{GN} = 0.36$,  
$\Lambda_{KKQT}/M_B=0.094$)
are also shown for comparison. It can be seen that there is little
difference for low $q^2$ between two kinds of calculations.
Say in other words, the inclusion of $\phi^+_B$ contribution can
be well approximated just by choosing a suitable value of the
parameter, $\omega_{KLS}$, $\omega_{GN}$ 
or $\Lambda_{KKQT}$. We found that the difference between 
the two kinds of calculations is about 3\% or less for $q^2 < 5$ GeV${}^2$.

For a reference we show the ratio of the contribution
from the $\phi_B^{\bar n}=(\phi_B^{-} - \phi_B^{+})/\sqrt{2}$ component
to that from  the all components in Table~\ref{tbl-kkqt3}.
The $\phi_B^{\bar n}$ component
contribution is found to be about 30\% or less.
\begin{table}[h]
\begin{tabular}{c|ccc}\hline
        & Gaussian & Exponential &KKQT   \\\hline
\rule{0mm}{5mm}
$F_{\phi^{\bar n}}^{B\pi}(0)/F_{total}^{B\pi}(0)$&0.22 &0.20&0.29\\\hline
\end{tabular}
\caption{Ratio of the sub-leading contribution to the total one in
$F^{B\pi}(0)$}
\label{tbl-kkqt3}
\end{table}
\section{Heavy to heavy form factors}
In this section we investigate the heavy-to-heavy
form factors in the fast recoil region, concentrating on the
$B\to D$ transition. We shall determine the parameters of
the $D$ meson distribution amplitude.
The $B\to D$ transition form factors are defined by the matrix elements,
\begin{equation}
\langle D (P_2)|{\bar b}(0)\gamma_\mu c(0)|B(P_1)\rangle
=\sqrt{m_Bm_D}\left[\xi_+(\eta)(v_1+v_2)_\mu+
\xi_-(\eta)(v_1-v_2)_\mu\right]\;,
\nonumber\\
\end{equation}
where $\eta = P_1\cdot P_2/(m_B m_D)$.
The lowest-order diagrams for the $B\to D$ form factors are
similar to Fig.~\ref{Bpiform1} replacing $u$ and $\pi$ by 
$c$ and $D$, respectively.
The leading-order formulae have been derived in \cite{TLS2}:
\begin{eqnarray}
\xi_+&=&16\pi C_F\sqrt{r}m_B^2\int dx_1dx_2 \int b_1db_1
b_2db_2\phi_B(x_1,b_1)\phi_D(x_2)
\nonumber\\
& &\times\left[E^D(t^{(1)}) h^D(x_1,x_2,b_1,b_2)
+rE^D(t^{(2)}) h^D(x_2,x_1,b_2,b_1)\right]\;,
\label{xip}\\
\xi_{-}&=&0\;,
\label{xim}
\end{eqnarray}
where the color factor $C_F=4/3$ and $r\equiv
m_D/m_B$. The functions $E^D(t)$ and
$h^D(x_1,x_2,b_1,b_2)$ are defined as
\begin{eqnarray}
E^D(t)&=&\alpha_s(t)\exp[-S_B(t)-S_D(t)]\;, \\
h^D(x_1,x_2,b_1,b_2)&=&K_{0}\left(\sqrt{x_1x_2r\eta^+}m_Bb_1\right)
S_t(x_2)
\nonumber \\
& &\times \left[\theta(b_1-b_2)K_0\left(\sqrt{x_2r\eta^+}m_B
b_1\right)I_0\left(\sqrt{x_2r\eta^+}m_Bb_3\right)\right.
\nonumber \\
&
&\left.+\theta(b_2-b_1)K_0\left(\sqrt{x_2r\eta^+}m_Bb_2\right)
I_0\left(\sqrt{x_2r\eta^+}m_Bb_1\right)\right]\;, \label{dh2}
\end{eqnarray}
where $\eta^+ = \eta + \sqrt{\eta^2 -1}$.
The definitions of the hard scales $t^{(1,2)}$ are as follows,
\begin{eqnarray}
t^{(1)}&=&{\rm max}(\sqrt{x_2r\eta^+}m_B,1/b_1,1/b_2)\;,
\nonumber\\
t^{(2)}&=&{\rm max}(\sqrt{x_1r\eta^+}m_B,1/b_1,1/b_2)\;.
\label{hat}
\end{eqnarray}
For numerical estimation, we use the model of $D$
meson distribution amplitude adopted in \cite{TLS2},
\begin{eqnarray}
\phi_D(x)=N_D x(1-x)[1+C_D(1-2x)]
\;, \label{ourDwv}
\end{eqnarray}
where $C_D$ is the $D$ meson distribution amplitude parameter. The normalization
constant $N_D$ is found to be $3f_D/\sqrt{2N_c}$
by using the relation
\begin{eqnarray}
\int dx\phi_D(x)=\frac{f_D}{2\sqrt{2N_c}}\;.
\end{eqnarray}
\subsection{Numerical Results}
Here we investigate the distribution amplitude dependence of
the $B \rightarrow D$ form factor. The inputs for $B$ meson
are same as the case of $B\rightarrow \pi$ form factor.
The $D$ meson distribution amplitude has
only one parameter $C_D$. We take $f_D= 240$ MeV, and 
other parameters are same as the $B\rightarrow \pi$ case
except for the threshold resummation parameter $c$ which 
is taken to be 0.35 in $B \rightarrow D$ transition\cite{TLS2}.
The $D$ meson distribution amplitude (\ref{ourDwv}) is decomposed into 
two parts;
\begin{equation}
\phi_D(x)= \phi_D^0(x) + C_D \phi_D^1(x)\; 
\end{equation} 
where $\phi_D^0(x)=N_D x(1-x)$ and $\phi_D^1(x)=N_D x(1-x)(1-2x)$.
The contributions from $\phi_D^0$ and $\phi_D^1$ at $\eta=1.58$ 
(near maximal recoil)
are shown in Table~\ref{xival} for 3 types of the $B$ meson distribution 
amplitudes. It can be seen that the value of $\xi_+$ varies 
about 4\% under the 10 \% change of $C_D$. 
We fix $C_D$ to be 1.5 so that the value of $\xi_+$ 
agrees with the experimental data, $\xi_+(1.58)\simeq 0.6$.
The $\eta$ dependence of $\xi_+$ is shown 
in Fig.~9 for $C_D=1.5$.
The results shows that the $B$ meson distribution amplitude 
dependence of $\xi_+$ is less than 5\%.
\begin{table}[h]
\begin{tabular}{c|ccc}\hline
contribution & Gaussian & Exponential & KKQT \\\hline
$\phi_D^0$&0.331&0.360&0.334\\ 
$\phi_D^1$&0.167&0.167&0.170\\
total ($C_D=1.5$)&0.582&0.610&0.589  \\\hline
\end{tabular} 
\caption{Contribution to $\xi_+$ from $\phi_D^0$ and $\phi_D^1$} 
\label{xival}
\end{table} 
\begin{figure}[h]
\resizebox{7cm}{!}{\includegraphics{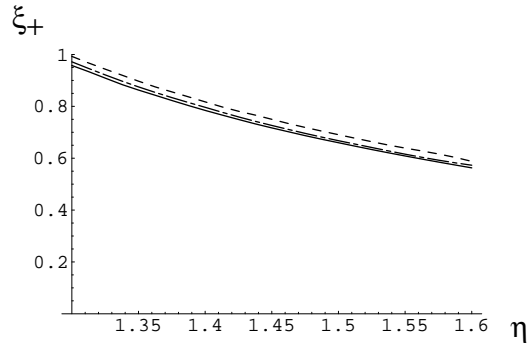}}
\caption{The $B\to D$ form factor $\xi_+$
as a function of $\eta= v_B\cdot v_D$. The results by Gaussian,
exponential and KKQT type B distribution amplitude are 
shown in solid, dot and dot-dashed line, respectively.
}
\label{xiplsfig} 
\end{figure}
\subsection{Another component of $B$ distribution amplitudes}
Following Sec. \ref{anoB} we investigate the contributions 
from the another component of the $B$ distribution amplitude
in the $B\rightarrow D$ form factor. The contribution from 
$\phi_B^n$ and $\phi_B^{\bar n}$ are given as
\begin{eqnarray}
\xi^{n}_{+} &=&
 -16\pi C_F\sqrt{r}m_B^2
\int dx_{1}dx_{2} \int b_1d b_1 b_2d b_2
\phi_B^{n}(x_1,b_1)\phi_{D}(x_2,b_2)
\nonumber \\
& &\times E^D(t^{(2)}) 
\;r h(x_2,x_1,b_2,b_1)
\;,\\
\xi^{\bar n}_{+} &=&
 -16\pi C_F\sqrt{r}m_B^2
\int d x_{1}d x_{2}\int b_1d b_1 b_2d b_2
\phi_B^{\bar n}(x_1,b_1)\phi_{D}(x_2,b_2)
\nonumber \\
& &\times  E^D(t^{(1)}) 
\;h(x_1,x_2,b_1,b_2)
\;,\\
\xi^{n}_{-} &=& \xi^{\bar n}_{-} = 0\;. 
\end{eqnarray}
The sum of contributions
from $\phi_B$, $\phi^{n}_B$ and $\phi^{\bar n}_B$
vanishes if $\phi_B = \phi^{n}_B = \phi^{\bar n}_B$ as in the 
case of $B\rightarrow \pi$.
We have investigated the contributions from $\phi_B^{\bar n}$ components
under the choice of $f=-\phi_B^+$ and
$\omega_{KLS} = 0.45$, $\omega_{GN} = 0.42$, $\Lambda_{KKQT}/m_B = 0.12$, 
which is obtained in $B\rightarrow \pi$ analysis.
The results of $\xi_+(1.58)$  with both $\phi_B^+$ and $\phi_B^-$ contributions
are shown in Table~\ref{xival2}.
It can be found that $C_D \simeq 0.6$ gives $\xi_+(1.58)\simeq  0.6$.
The difference due to the choice of the $B$ distribution amplitude 
becomes about 16\% or less here.
The $\phi_B^{\bar n}$ component contribution is not numerically 
sub-leading in the case of KKQT type distribution amplitude.

\begin{figure}[h]
\resizebox{5.5cm}{!}{\includegraphics{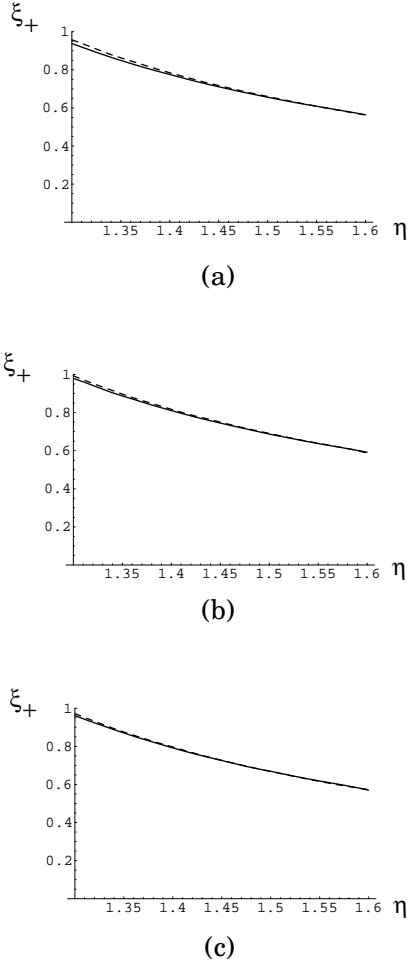}}
\caption{The $B\to D$ form factors $\xi_+$ as functions of $\eta$
for Gaussian type ($C_D=0.74$) (a), exponential type ($C_D=0.77$) (b) and
KKQT type ($C_D=0.40$) (c) distribution amplitudes.
The results with the leading contribution only
with a suitable choice of the parameter
are shown in dashed lines, while
those with both contributions are shown in solid lines.
}
\label{xi-sub}
\end{figure}

The $\xi_+$ value at $\eta=1.58$ changes 5$\sim$8\% by the inclusion 
of $\phi_B^{\bar n}$ contribution as seen by comparing the results 
given in Tables~\ref{xival} and \ref{xival2}. 
If a suitable value of $C_D$ is taken for each $B$ distribution amplitudes, 
we can reduce the difference. The suitable choice is 
$C_D=0.74$, 0.77 and 0.40 for 
Gaussian, exponential and KKQT, respectively.
The $\eta$ dependence of $\xi_+$ with both contributions with the 
suitable value of $C_D$ is
shown in Fig.\ref{xi-sub}. The results with the leading contribution only
($\omega_{KLS} = 0.38$, $\omega_{GN} = 0.36$, 
$\Lambda_{KKQT}/M_B=0.094$, $C_D=1.5$)
are also shown for comparison. It can be seen that there is little
difference for $1.3 \le \eta \le 1.58$ between two kinds of calculations.
The inclusion of the sub-leading contribution can
be well approximated just by choosing a suitable value of the
parameter, $C_D$ as in the case of $B\rightarrow \pi$. 
We found that the difference between 
the two kinds of calculations is about 2\% or less.
\begin{table}[h]
\begin{tabular}{c|ccc}\hline
contribution & Gaussian & Exponential & KKQT \\\hline
$\xi_+ (\phi_D^0)$&0.246&0.277&0.220\\
$\xi_+^{\bar n} (\phi_D^0)$&0.176&0.165&0.265\\
$\xi_+ (\phi_D^1)$&0.124&0.130&0.111\\
$\xi_+^{\bar n} (\phi_D^1)$&0.093&0.089&0.153\\
total ($C_D=0.6$)&0.552&0.573&0.643\\\hline
\end{tabular}
\caption{Contribution to $\xi_+$ and $\xi_+^{\bar n}$ 
from $\phi_D^0$ and $\phi_D^1$}
\label{xival2}
\end{table}
\section{$B\rightarrow D\pi$}
The decay rates of $B\to D\pi$ is given as 
\begin{equation}
\Gamma_i=\frac{1}{128\pi}G_F^2|V_{cb}|^2|V_{ud}|^2\frac{m_B^3}{r}
|{\cal M}_i|^2\;,
\label{dr}
\end{equation}
where $r \equiv m_D/m_B$. The indices,
$i=1$, 2, and 3, denote the modes 
${\bar B}^0\to D^+\pi^-$, 
${\bar B}^0\to D^0\pi^0$ and 
$B^- \to D^0\pi^-$
 respectively. The decay
amplitudes ${\cal M}_i$ are written as\cite{KKLL}
\begin{eqnarray}
{\cal M}_1&=&f_\pi \xi_{\rm ext} + f_B\xi_{\rm exc}+{\cal M}_{\rm ext}
+{\cal M}_{\rm exc}\;,
\label{M1}\\
{\cal M}_2&=& -\frac{1}{\sqrt{2}}[ f_D\xi_{\rm int} - f_B\xi_{\rm exc}
      +{\cal M}_{\rm int} -{\cal M}_{\rm exc} ]
\label{M2}\\
{\cal M}_3&=&f_\pi\xi_{\rm ext} +
f_D\xi_{\rm int}+{\cal M}_{\rm ext}+{\cal M}_{\rm int}\;.
\label{M3}
\end{eqnarray}
The factor
$\xi_{\rm ext}$ denotes the factorizable external
$W$-emission contributions. The factors $\xi_{\rm int}$ and
$\xi_{\rm exc}$ represent the factorizable internal $W$-emission and
$W$-exchange contributions, respectively.
The amplitudes ${\cal M}_{\rm ext}$,
${\cal M}_{\rm int}$, and ${\cal M}_{\rm exc}$ are the
non-factorizable external $W$-emission, internal $W$-emission, and
$W$-exchange contributions, respectively.
The factor $\xi_{\rm ext}$ ($\xi_{\rm int}$) is obtained by   
the convolution between the Wilson coefficients and 
$B\to D$ ($\pi$) form factor. The leading 
formulae of these expressions are given in \cite{KKLL}. They are 
summarized with the $\phi_B^n$ and $\phi_B^{\bar n}$ contributions 
in the Appendix C. 

Let us first show the leading order calculation for each 
$B$ distribution amplitude without $n$ and $\bar n$ contributions. 
The parameters are the same in the cases of the form factor calculations. 
($\omega_{KLS}=0.38$, $\omega_{GN}=0.36$, $\Lambda_{KKQT}/M_B=0.094$ 
and $C_D=1.5$)
The result is shown in Table~\ref{dpileadtbl}.
Our result of Gaussian case is slightly different from that 
given in \cite{KKLL}.
It is partly due to the choice of the parameters and partly 
due to the change of anomalous dimension adopted 
in the Sudakov factor\cite{LL04}. We should look at the 
ratios between branching ratios rather than 
the magnitudes of the branching ratios since there is uncertainty in 
the decay constants of heavy mesons which gives overall normalization 
of the distribution amplitudes. BR($D^+\pi^-$) is slightly 
larger than the experimental data, while BR($D^0\pi^0$) is 
slightly smaller than that.
\begin{table}[h]
\begin{tabular}{c|llll}\hline
decay mode & Gaussian & Exponential & KKQT & Exp.\\\hline
$D^0\pi^-$& 5.3 (1.0)& 5.2 (1.0)& 5.2 (1.0)& $4.98 \pm 0.29$ (1.0)\\
$D^+\pi^-$& 3.2 (0.60)& 3.7 (0.71)& 3.3 (0.63)& 
$2.76\pm 0.25$ ($0.55\pm 0.06$)\\
$D^0\pi^0$& 0.18 (0.034)& 0.11 (0.021)& 0.20 (0.039) &
 $0.291 \pm 0.028$ ($0.058\pm 0.007$) \\\hline
\end{tabular}
\caption{%
The branching ratios of $B\rightarrow D\pi$ decay modes in the unit of $10^{-3}$. 
The number in the parenthesis is the ratio to BR($D^0\pi^-$). 
The experimental data is from \cite{PDG}.  
} 
\label{dpileadtbl}
\end{table}

\begin{figure}[h]
\resizebox{6cm}{!}{%
\includegraphics{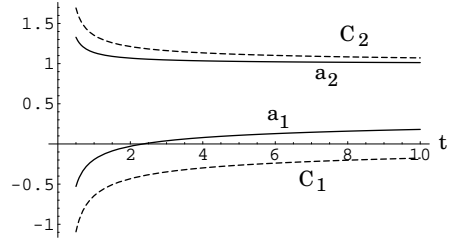}
}
\caption{Scale dependence of the Wilson coefficients.}
\label{Wilson}
\end{figure}

There is a cancellation between the Wilson coefficients $C_1$ and $C_2$ in 
the evaluation of   $a_1(t) = C_1(t) + C_2(t)/N_C$ 
which enters in $\xi_{\rm int}$ and $\xi_{\rm exc}$
as can be seen in Fig.{\ref{Wilson}}. $a_1(t)$ almost vanishes around 
$t \simeq M_B/2$. The contribution from $\xi_{\rm int}$ is numerically 
significant in the ${\bar B}^0\to D^0\pi^0$ decay amplitude. (The 
contribution from $\xi_{\rm exc}$ is negligible.) 
For reference, we show 
how the branching ratios change if we adopt the fixed scale for 
the evaluation of the Wilson coefficients and $\alpha_s$ 
in Table~\ref{dpiscale}. The result shows that the choice of 
the scale $t$ can give large uncertainty in $B\rightarrow D\pi$.
\begin{table}[h]
\begin{tabular}{lc|lll}
&decay mode & Gaussian & Exponential & KKQT \\\hline
fixed $t=M_B/2$
&$D^0\pi^-$& 6.8 (1.0)& 6.5 (1.0)& 6.7 (1.0)\\
&$D^+\pi^-$& 2.6. (0.38)& 2.8 (0.43)&  2.6(0.39) \\
&$D^0\pi^0$& 0.44 (0.065)& 0.34 (0.052)& 0.46 (0.069) \\\hline
fixed $t=M_B$
&$D^0\pi^-$& 7.4 (1.0)& 7.1 (1.0)& 7.3(1.0)\\
&$D^+\pi^-$& 2.2 (0.30)&  2.4 (0.34)& 2.2 (0.30) \\
&$D^0\pi^0$& 0.66 (0.089)& 0.53 (0.075)& 0.69 (0.095) \\\hline
fixed $t=2M_B$
&$D^0\pi^-$& 8.0 (1.0)&  7.6 (1.0)& 7.9 (1.0)\\
&$D^+\pi^-$& 2.0 (0.25)&  2.2 (0.29)& 2.0 (0.25) \\
&$D^0\pi^0$& 0.87 (0.11)& 0.72 (0.95)& 0.89 (0.11) \\\hline
\end{tabular} 
\caption{%
The branching ratios of $B\rightarrow D\pi$ decay modes 
for fixed RGE scale in the unit of $10^{-3}$.
The number in the parenthesis is the ratio to BR($D^0\pi^-$).
}
\label{dpiscale}
\end{table}

Let us estimate the $\phi_B^n$ and $\phi_B^{\bar n}$ contributions.
The formulae of $B\rightarrow D\pi$ amplitudes in PQCD calculation are 
obtained under the following choice of the light quark 
momenta in $B$ meson\cite{KKLL};
\begin{eqnarray}
  k_1 &=& \frac{x_1M_B}{\sqrt{2}}(1,0,0_T) + k_{1T}
\text{\ \  for\ \  $\xi_{\rm int}$, $M_{\rm int}$,} \\
  k_1 &=& \frac{x_1M_B}{\sqrt{2}}(0,1,0_T) + k_{1T}
\text{\ \  for\ \  others.}
\end{eqnarray}
Then we should take the leading $B$ meson distribution amplitude as 
$\phi_B^+$ in $\xi_{\rm int}$ and $M_{\rm int}$, while 
$\phi_B^-$ in others. 
(Remind the discussion given in Sec.~\ref{anoB}.)
The parameters of the distribution amplitudes are 
taken as $\omega_{KLS}=0.45$, $\omega_{GN}=0.42$, $\Lambda_{KKQT}/M_B=0.12$
and $C_D=0.6$. 
The ratios of the $\phi_B^n$ and $\phi_B^{\bar n}$ contribution to 
the total one in each component of the decay amplitude are shown in 
Table~\ref{dpiamptbl}. Re$M_{exc}$ receives large contributions from $\phi_B^n$ 
and $\phi_B^{\bar n}$. But its magnitude is far smaller than those of 
$\xi_{ext}$, $\xi_{int}$ and $M_{int}$, so that 
the effect is not significant in the total amplitudes 
Eqs.(\ref{M1})-(\ref{M3}). The $\phi_B^n$ and 
$\phi_B^{\bar n}$ contribution to $M_{ext}$ vanishes since $x_3$ and 
$(1-x_3)$ terms cancels in $M_{ext}^{\bar n}$.
The branching ratios calculated with this set of parameters are 
given in Table~\ref{dpileadtbl2}. BR($D^+\pi^-$) gets lower and 
approaches the experimental value from the view point of the ratio. 
BR($D^0\pi^0$) becomes larger except for KKQT type case, 
which is a good tendency to realize the experimental values.
The Gaussian type distribution amplitude becomes the 
best candidate here.
\begin{table}[h]
\begin{tabular}{c|rrr}\hline
 & Gaussian & Exponential & KKQT \\\hline
$\xi_{ext}$&0.42&0.37&0.55 \\
$\xi_{int}$&0.16&0.13&0.23 \\
$\text{Re}M_{ext}$&0.0&0.0&0.0 \\
$\text{Im}M_{ext}$&0.0&0.0&0.0 \\
$\text{Re}M_{int}$&0.37&0.38&0.50 \\
$\text{Im}M_{int}$&$-0.08$&$-0.04$&$-0.05$\\
$\text{Re}M_{exc}$&$-1.13$&$-0.85$&$-1.25$\\
$\text{Im}M_{exc}$&0.16&0.20&0.25 \\\hline
\end{tabular}
\caption{%
The ratios, ($\phi_B^n$ and $\phi_B^{\bar n}$ contribution)/(total), 
in each amplitude with $C_D=0.6$. 
} 
\label{dpiamptbl}
\end{table}
\begin{table}[h]
\begin{tabular}{c|lll}\hline
decay mode & Gaussian & Exponential & KKQT \\\hline
$D^0\pi^-$& 6.5 (1.0)&  6.2 (1.0)& 8.4 (1.0)\\
$D^+\pi^-$& 3.2 (0.50)& 3.5 (0.57)& 4.5 (0.54)\\
$D^0\pi^0$& 0.25 (0.038)& 0.17 (0.027)& 0.27 (0.032) \\\hline
\end{tabular}
\caption{%
The branching ratios of each decay modes in the unit of $10^{-3}$ 
with $C_D=0.6$.
The number in the parenthesis is the ratio to BR($D^0\pi^-$).
}
\label{dpileadtbl2}
\end{table}

Next we take the parameters as $C_D= 0.74$, 0.77 and 0.4 for 
Gaussian, exponential and KKQT type distribution amplitudes, respectively 
as done in the case of the $B\rightarrow D$ form factor calculation.
The ratios of the decay amplitude with $\phi_B^n$ and $\phi_B^{\bar n}$ 
contributions to that of the leading calculation adopted 
in obtaining Table~\ref{dpileadtbl} are shown in Table~\ref{dpiamptbl2}. It can be found that 
the leading calculation gives a good approximation with the uncertainty 
about 20\%. The branching ratios in this calculation are given in 
Table~\ref{dpileadtbl3}. This result also shows a good tendency to approach the 
experimental value in comparison with the result of the 
leading calculation given in Table~\ref{dpileadtbl}. The KKQT type distribution 
amplitude becomes the best candidate in this case.
\begin{table}[htc]
\begin{tabular}{l|lll}\hline
 & Gaussian & Exponential & KKQT \\\hline
$A(D^0\pi^-)$&1.13 ($4.3^\circ$)&1.13 ($2.8^\circ$)&1.20 ($5.5^\circ$)\\
$A(D^+\pi^-)$&1.06 ($-0.6^\circ$)&1.04 ($-1.1^\circ$)&1.08 ($-0.5^\circ$)\\
$A(D^0\pi^0)$&1.13 ($22^\circ$)&1.18 ($27^\circ$)&1.22 ($29^\circ$)\\\hline
\end{tabular}
\caption{%
The ratios of the decay amplitudes, 
(calculation
with $n$ and $\bar n$ contribution)/ 
(leading calculation given in Table~XIII)
in each decay mode 
with $C_D= 0.74$, 0.77 and 0.4 for Gaussian, exponential and
KKQT type distribution amplitudes, respectively. The number 
in the parenthesis is the phase.
}
\label{dpiamptbl2}
\end{table}
\begin{table}[h]
\begin{tabular}{c|lll}\hline
decay mode & Gaussian & Exponential & KKQT \\\hline
$D^0\pi^-$& 6.9 (1.0)&  6.7 (1.0)& 7.7 (1.0)\\
$D^+\pi^-$& 3.6 (0.52)& 4.0 (0.60)& 3.8 (0.50)\\
$D^0\pi^0$& 0.23 (0.034)& 0.15 (0.022)& 0.30 (0.040) \\\hline
\end{tabular}
\caption{%
The branching ratios of each decay modes in the unit of $10^{-3}$
with $C_D= 0.74$, 0.77 and 0.4 for Gaussian, exponential and 
KKQT type distribution amplitudes, respectively.
The number in the parenthesis is the ratio to BR($D^0\pi^-$).
}
\label{dpileadtbl3}
\end{table}
\section{Summary and Discussions}
We have analyzed the uncertainty in the PQCD calculations of 
$B\rightarrow \pi$, $B\rightarrow D$ form factors and 
$B\rightarrow D\pi$ decay rates. The sources of uncertainty in 
$B\rightarrow \pi \; (D)$ form factors are summarized 
in Table ~ref{tbl-errs}. The uncertainty in the perturbative 
hard part is less than 10\%. The major source of uncertainty comes 
from the meson distribution amplitudes. 
The meson distribution 
amplitude is a non-perturbative quantity, so that we need 
a model or a non-perturbative method to evaluate it. 
The leading PQCD results varies 10$\sim$ 30 \% by changing the 
parameters in the meson distribution amplitudes.
The uncertainty from the RGE scale choice is small in the
form factor, while it is large due to subtle cancellation
between Wilson coefficients in $B\rightarrow D\pi$.

\begin{table}[h]
\begin{tabular}{c|c|c} \hline
mode&source   &  uncertainty \\\hline
$B\rightarrow \pi$ & $\omega_B$, $m_0$ &  30\% \\
&$a_2 , \dots , a_{2t}$  & 10\%\\
&$b$ dependence of pion & 10\%\\
&$f_B$, $f_\pi$ &  normalization \\
&evolution effects & 10\%\\
&$\Lambda_{\rm QCD}$&  3\%\\
&choice of the hard scale &  10\%\\
&$x_1$ terms & 4\%\\
&another $B$ distribution amplitude & 
20$\sim$30 \% ${}^{*)}$
\\\hline
$B\rightarrow D$ & $C_D$ & 4\%\\
&another $B$ distribution amplitude & 
40$\sim$60 \% ${}^{*)}$
\\\hline
\end{tabular}
\caption{The uncertainty from each source. ${}^{*)}$
As for the uncertainty from the another $B$ distribution amplitude, 
the effects of the inclusion of another $B$ distribution amplitude 
can be well approximated by a single $B$ distribution amplitude 
with a suitable value of the parameter as shown in Figs. 8 and 10.
}
\label{tbl-errs}
\end{table}

Here we have tried three kinds of the $B$ meson distribution amplitudes.
Two of them are models and one is derived from the equations of motion 
under the neglection of 3-parton contributions. It is surprising that
the three types of $B$ meson distribution amplitudes give almost same 
PQCD results of $B\rightarrow \pi \; (D)$ form factors by suitably 
choosing their parameters although the functional forms of them 
are rather different with one another. The non-factorizable contributions 
in non-leptonic $B$ decays can be of help to discriminate the $B$ meson 
distribution amplitudes. 

The formally sub-leading component of the $B$ meson distribution amplitude 
gives significant contributions to $B$ decays. This component is neglected 
in many of the previous PQCD calculations. But the leading $B$ meson distribution 
amplitude alone can give a good approximation if we suitably choose the 
parameters. The difference can be reduced to be a few \% for the form factors, 
and about 20 \% for $B\rightarrow D\pi$ amplitudes with a suitable parameter 
choice. So the results of the previous PQCD studies are still useful. 
\begin{acknowledgments}
The author would like to thank Prof. H-n. Li, Prof. A.I. Sanda and
other members of PQCD working group for fruitful discussions and
encouragement.
The author would like to show gratitude for the Summer Institute 2005 at
Fuji-Yoshida, where a part of this work was done.
\end{acknowledgments}
\appendix
\section{Approximation formulae of $B\rightarrow\pi$ form factors}
In the case of exponential type $B$ meson distribution amplitude
the $\omega_{GN}$ dependence can be well approximated 
by the following formulae for $0.26 \le \omega_{GN} \le 0.46$;
\begin{eqnarray}
F^{A0}(\omega_{GN})&=& 0.0597 - 0.184\,(\omega_{GN} -0.36 )  + 0.459\,(\omega_{GN} -0.36 )^2
        - 1.04\,(\omega_{GN} -0.36 )^3\;, \nonumber\\
F^{A2}(\omega_{GN})&=& 0.0780 - 0.235\,(\omega_{GN} -0.36 )  + 0.537\,(\omega_{GN} -0.36 )^2
         - 1.06\,(\omega_{GN} -0.36 )^3\;, \nonumber\\
F^{A4}(\omega_{GN})&=& 0.0702 - 0.212\,(\omega_{GN} -0.36 )  + 0.485\,(\omega_{GN} -0.36 )^2
         - 0.701\,(\omega_{GN} -0.36 )^3\;, \nonumber\\
F^{P0}(\omega_{GN})&=& 0.507 - 2.38\,(\omega_{GN} -0.36 )  +  8.60\,(\omega_{GN} -0.36 )^2
   - 24.7\,(\omega_{GN} -0.36 )^3\;,  \nonumber\\
F^{P2}(\omega_{GN})&=& 0.144 - 0.537\,(\omega_{GN} -0.36 )  + 1.52,(\omega_{GN} -0.36 )^2
   - 3.55\,(\omega_{GN} -0.36 )^3\;, \nonumber\\
F^{P4}(\omega_{GN})&=& 0.0748 - 0.259\,(\omega_{GN} -0.36 )  + 0.649\,(\omega_{GN} -0.36 )^2
   - 0.994\,(\omega_{GN} -0.36 )^3\;, \nonumber\\
F^{T0}(\omega_{GN})&=& 0.0986 - 0.304\,(\omega_{GN} -0.36 )  + 0.711\,(\omega_{GN} -0.36 )^2
   - 1.37\,(\omega_{GN} -0.36 )^3\;, \nonumber\\
F^{T2}(\omega_{GN})&=& 0.394 - 1.23\,(\omega_{GN} -0.36 )  + 2.84\,(\omega_{GN} -0.36 )^2
   - 5.09\,(\omega_{GN} -0.36 )^3\; .
\end{eqnarray}

In the case of KKQT type $B$ meson distribution amplitude
the $\Lambda_{KKQT}$ dependence can be well approximated
by the following formulae for $0.074 \le \Lambda_{KKQT}/M_B \le 0.114$;
\begin{eqnarray}
F^{A0}(\Lambda_{KKQT})&=& 0.0604 - 0.662\,(\Lambda_{KKQT}/M_B -0.094 )  + 5.55\,(\Lambda_{KKQT}/M_B -0.094 )^2
        - 56.8\,(\Lambda_{KKQT}/M_B -0.094 )^3\;, \nonumber\\
F^{A2}(\Lambda_{KKQT})&=& 0.0876 - 0.946\,(\Lambda_{KKQT}/M_B -0.094 )  + 6.44\,(\Lambda_{KKQT}/M_B -0.094 )^2
         - 56.2\,(\Lambda_{KKQT}/M_B -0.094 )^3\;, \nonumber\\
F^{A4}(\Lambda_{KKQT})&=& 0.0817 - 0.887\,(\Lambda_{KKQT}/M_B -0.094 )  + 5.61\,(\Lambda_{KKQT}/M_B -0.094 )^2
         - 27.3\,(\Lambda_{KKQT}/M_B -0.094 )^3\;, \nonumber\\
F^{P0}(\Lambda_{KKQT})&=& 0.450 - 7.45\,(\Lambda_{KKQT}/M_B -0.094 )  +  90.9\,(\Lambda_{KKQT}/M_B -0.094 )^2
   - 1160\,(\Lambda_{KKQT}/M_B -0.094 )^3\;,  \nonumber\\
F^{P2}(\Lambda_{KKQT})&=& 0.157 - 2.12\,(\Lambda_{KKQT}/M_B -0.094 )  + 19.3,(\Lambda_{KKQT}/M_B  -0.094 )^2
   - 203\,(\Lambda_{KKQT}/M_B -0.094 )^3\;, \nonumber\\
F^{P4}(\Lambda_{KKQT})&=& 0.0864 - 1.07\,(\Lambda_{KKQT}/M_B -0.094 )  + 7.57\,(\Lambda_{KKQT}/M_B  -0.094 )^2
   - 160\,(\Lambda_{KKQT}/M_B -0.094 )^3\;, \nonumber\\
F^{T0}(\Lambda_{KKQT})&=& 0.113 - 1.27\,(\Lambda_{KKQT}/M_B -0.094 )  + 8.19\,(\Lambda_{KKQT}/M_B  -0.094 )^2
   - 64.7\,(\Lambda_{KKQT}/M_B -0.094 )^3\;, \nonumber\\
F^{T2}(\Lambda_{KKQT})&=& 0.465 - 5.35\,(\Lambda_{KKQT}/M_B -0.094 )  + 33.5\,(\Lambda_{KKQT}/M_B -0.094 )^2
   - 200\,(\Lambda_{KKQT}/M_B -0.094 )^3\; .\nonumber\\
&&
\end{eqnarray}
\section{$\phi_B^-$ in Gaussian type distribution amplitude}
The equations of motion for $\phi_B^+$ and $\phi_B^-$
are given with the approximation of neglecting 3-parton
contributions as\cite{KKQT}
\begin{eqnarray}
 \phi_B^+(x) + x {\phi_B^-}'(x) &=& 0 \; ,\label{eqhqet1}\\
\left(x- \frac{2\bar \Lambda}{m_B}\right) \phi_B^+(x) +
x \phi_B^-(x) &=& 0  \;, \label{eqhqet2}
\end{eqnarray}
where $\bar \Lambda = m_B -m_b$ is the hadronic scale of HQET.
By solving Eq.(\ref{eqhqet1}) with $\phi_B^{-} = \phi_B^{KLS}$
we obtain Eq.(\ref{gaussmns}). As for Eq.(\ref{eqhqet2}) the
left hand side dons not necessary vanishes, but its value
is less than 10\% of $\phi_B^+(0)$ for $\bar\Lambda/m_B \simeq 0.1$.
\section{$B \to D\pi$ formuale}
The contributions to $\xi_{ext}$ are given as
\begin{eqnarray}
\xi_{\rm ext} &=&
 16\pi C_F\sqrt{r}m_B^2
\int_{0}^{1}d x_{1}d x_{2}\int_{0}^{1/\Lambda} b_1d b_1 b_2d b_2
\phi_B(x_1,b_1)\phi_{D}(x_2,b_2)
\nonumber \\
& &\times\alpha_s(t)a_2(t)\exp[-S_B(t)-S_{D}(t)]
\nonumber \\
& &\times\left[
h(x_1,x_2,b_1,b_2) + r h(x_2,x_1,b_2,b_1)
\right],\\
\xi_{\rm ext}^{\bar n} &=&
 16\pi C_F\sqrt{r}m_B^2
\int_{0}^{1}d x_{1}d x_{2}\int_{0}^{1/\Lambda} b_1d b_1 b_2d b_2
\phi_B^{\bar n}(x_1,b_1)\phi_{D}(x_2,b_2)
\nonumber \\
& &\times\alpha_s(t)a_2(t)\exp[-S_B(t)-S_{D}(t)]
\left[ -h(x_1,x_2,b_1,b_2) \right]
,\\
\xi_{\rm ext}^{n} &=&
 16\pi C_F\sqrt{r}m_B^2
\int_{0}^{1}d x_{1}d x_{2}\int_{0}^{1/\Lambda} b_1d b_1 b_2d b_2
\phi_B^{n}(x_1,b_1)\phi_{D}(x_2,b_2)
\nonumber \\
& &\times\alpha_s(t)a_2(t)\exp[-S_B(t)-S_{D}(t)] \\
&& \times
\left[ -r h(x_2,x_1,b_2,b_1)\right],
\end{eqnarray}
where
$a_2 = C_2 + C_1/N_c$, and $C_{1,2}$ are the Wilson coefficients.

The  contributions to $\xi_{\rm int}$ are given as,
\begin{eqnarray}
\xi_{\rm int}&=&16\pi C_F\sqrt{r}m_B^2
\int_0^1 dx_1dx_3\int_0^{1/\Lambda}b_1db_1b_3db_3
\phi_B(x_1,b_1) 
\nonumber \\
& &\times \alpha_s(t_{\rm int})a_1(t_{\rm int})
\exp[-S_B(t_{\rm int})-S_\pi(t_{\rm int})]
\nonumber \\
& &\times
\biggl[
\left[
(1+x_3) \phi_\pi(x_3)
+ r_0 ( 1- 2 x_3) (\phi_\pi^p(x_3) + \phi_\pi^t(x_3))
\right]
h(x_1,x_3(1-r^2),b_1,b_3)
\\
& &
\ \ \ + 2r_0 \phi_\pi^p(x_3) h(x_3,x_1(1-r^2),b_3,b_1)
\biggr]
\;,
\\
\xi_{\rm int}^{\bar n}&=&16\pi C_F\sqrt{r}m_B^2
\int_0^1 dx_1dx_3\int_0^{1/\Lambda}b_1db_1b_3db_3
\phi_B^{\bar n}(x_1,b_1) 
\nonumber \\
& &\times \alpha_s(t_{\rm int})a_1(t_{\rm int})
\exp[-S_B(t_{\rm int})-S_\pi(t_{\rm int})]
\nonumber \\
& &\times [
-x_3 \phi_\pi(x_3)
- r_0 (1 -x_3) (\phi_\pi^p(x_3) + \phi_\pi^t(x_3))
]
h(x_1,x_3(1-r^2),b_1,b_3) \;,
\nonumber \\
& &
 \ \ \  - 2r_0 \phi_\pi^p(x_3)
h(x_3,x_1(1-r^2),b_3,b_1)\Biggr]\;,
\label{int} \\\
\xi_{\rm int}^{n}&=&16\pi C_F\sqrt{r}m_B^2
\int_0^1 dx_1dx_3\int_0^{1/\Lambda}b_1db_1b_3db_3
\phi_B^{n}(x_1,b_1) 
\nonumber \\
& &\times \alpha_s(t_{\rm int})a_1(t_{\rm int})
\exp[-S_B(t_{\rm int})-S_\pi(t_{\rm int})]
\nonumber \\
& &\times [
- \phi_\pi(x_3)
+ r_0 x_3 (\phi_\pi^p(x_3) + \phi_\pi^t(x_3))
]
h(x_1,x_3(1-r^2),b_1,b_3)
\label{int2}
\end{eqnarray}
where $a_1 = C_1 + C_2/N_c$.

The form factor $\xi_{\rm exc}$ is written as
\begin{eqnarray}
\xi_{\rm exc}&=&16\pi C_F\sqrt{r}m_B^2
\int_0^1 dx_2dx_3\int_0^{1/\Lambda}b_2db_2b_3db_3
\phi_{D}(x_2,b_2)
\nonumber \\
& &\times\alpha_s(t_{\rm exc})a_1(t_{\rm exc})
\exp[-S_{D}(t_{\rm exc})-S_\pi(t_{\rm exc})]
\nonumber \\
& &\times
\Biggl[ 
-x_3\phi_\pi(x_3) h_a(x_2,x_3(1-r^2),b_2,b_3)
\nonumber \\
& &\ \ \ \  + x_2\phi_\pi(x_3)
h_a(x_3,x_2(1-r^2),b_3,b_2)
\Biggr]\;.
\label{exc}
\end{eqnarray}
The $B$ distribution amplitude does not enter in $\xi_{\rm exc}$, 
so there is no $\phi_B^{n(\bar n)}$ contribution here.

For the non-factorizable amplitudes,
their expressions are
\begin{eqnarray}
{\cal M}_{\rm ext}&=& 32\pi\sqrt{2N} C_F\sqrt{r}m_B^2
\int_0^1 [dx]\int_0^{1/\Lambda}
b_1 db_1 b_3 db_3
\phi_B(x_1,b_1)\phi_{D^{(*)}}(x_2,b_1)\phi_\pi(x_3)
\nonumber \\
& &\times \alpha_s(t_b)\frac{C_1(t_b)}{N}\exp[-S(t_b)|_{b_1=b_2}]
\nonumber \\
& &\times \Biggl[
x_3 h^{(1)}_b(x_i,b_i)
-(1-x_3 +x_2)h^{(2)}_b(x_i,b_i) \Biggr]\;,
\\
{\cal M}^{\bar n}_{\rm ext}&=& 32\pi\sqrt{2N} C_F\sqrt{r}m_B^2
\int_0^1 [dx]\int_0^{1/\Lambda}
b_1 db_1 b_3 db_3
\phi^{\bar n}_B(x_1,b_1)\phi_{D^{(*)}}(x_2,b_1)\phi_\pi(x_3)
\nonumber \\
& &\times \alpha_s(t_b)\frac{C_1(t_b)}{N}\exp[-S(t_b)|_{b_1=b_2}]
\nonumber \\
& &\times \Biggl[
-x_3 h^{(1)}_b(x_i,b_i)
+(1-x_3)h^{(2)}_b(x_i,b_i)
\Biggr]\;,
\\%
{\cal M}^{n}_{\rm ext}&=& 32\pi\sqrt{2N} C_F\sqrt{r}m_B^2
\int_0^1 [dx]\int_0^{1/\Lambda}
b_1 db_1 b_3 db_3
\phi_B^{n}(x_1,b_1)\phi_{D^{(*)}}(x_2,b_1)\phi_\pi(x_3)
\nonumber \\
& &\times \alpha_s(t_b)\frac{C_1(t_b)}{N}\exp[-S(t_b)|_{b_1=b_2}]
\nonumber \\
& &\times \Biggl[
x_2 h^{(2)}_b(x_i,b_i) \Biggr]\;.
\end{eqnarray}
\begin{eqnarray}
{\cal M}_{\rm int}&=& 32\pi\sqrt{2N} C_F\sqrt{r}m_B^2
\int_0^1 [dx]\int_0^{1/\Lambda}b_1 db_1 b_2 db_2
\phi_B(x_1,b_1)\phi_{D^{(*)}}(x_2,b_2)\phi_\pi(x_3)
\nonumber \\
& &\times \alpha_s(t_d)\frac{C_2(t_d)}{N}\exp[-S(t_d)|_{b_3=b_1}]
\nonumber \\
& &\times
\Biggl[
  (-x_2-x_3)
h^{(1)}_d(x_i,b_i)
+ (1-x_2)
h^{(2)}_d(x_i,b_i)
\Biggr]\;,
\\
{\cal M}^{\bar n}_{\rm int}&=& 32\pi\sqrt{2N} C_F\sqrt{r}m_B^2
\int_0^1 [dx]\int_0^{1/\Lambda}b_1 db_1 b_2 db_2
\phi^{\bar n}_B(x_1,b_1)\phi_{D^{(*)}}(x_2,b_2)\phi_\pi(x_3)
\nonumber \\
& &\times \alpha_s(t_d)\frac{C_2(t_d)}{N}\exp[-S(t_d)|_{b_3=b_1}]
\nonumber \\
& &\times
\Biggl[
  x_3
h^{(1)}_d(x_i,b_i)
\Biggr]\;,\\
{\cal M}^{n}_{\rm int}&=& 32\pi\sqrt{2N} C_F\sqrt{r}m_B^2
\int_0^1 [dx]\int_0^{1/\Lambda}b_1 db_1 b_2 db_2
\phi^{n}_B(x_1,b_1)\phi_{D^{(*)}}(x_2,b_2)\phi_\pi(x_3)
\nonumber \\
& &\times \alpha_s(t_d)\frac{C_2(t_d)}{N}\exp[-S(t_d)|_{b_3=b_1}]
\nonumber \\
& &\times
\Biggl[
  x_2 h^{(1)}_d(x_i,b_i)
- (1-x_2)
h^{(2)}_d(x_i,b_i)
\Biggr]\;,
\end{eqnarray}
\begin{eqnarray}
{\cal M}_{\rm exc}&=& 32 \pi\sqrt{2N} C_F\sqrt{r}m_B^2
\int_0^1 [dx]\int_0^{1/\Lambda}b_1 db_1 b_2 db_2
\phi_B(x_1,b_1)\phi_{D}(x_2,b_2) \phi_\pi(x_3)
\nonumber \\
& &\times \alpha_s(t_f)\frac{C_2(t_f)}{N}\exp[-S(t_f)|_{b_3=b_2}]
\nonumber \\
& &\times \Biggl[
x_3
h^{(1)}_f(x_i,b_i)
- x_2  h^{(2)}_f(x_i,b_i)
\Biggr]\;,
\\
{\cal M}^{\bar n}_{\rm exc}&=& 32 \pi\sqrt{2N} C_F\sqrt{r}m_B^2
\int_0^1 [dx]\int_0^{1/\Lambda}b_1 db_1 b_2 db_2
\phi^{\bar n}_B(x_1,b_1)\phi_{D}(x_2,b_2) \phi_\pi(x_3)
\nonumber \\
& &\times \alpha_s(t_f)\frac{C_2(t_f)}{N}\exp[-S(t_f)|_{b_3=b_2}]
\nonumber \\
& &\times \Biggl[
-x_3
h^{(1)}_f(x_i,b_i)
\Biggr]\;,
\\
{\cal M}^{n}_{\rm exc}&=& 32 \pi\sqrt{2N} C_F\sqrt{r}m_B^2
\int_0^1 [dx]\int_0^{1/\Lambda}b_1 db_1 b_2 db_2
\phi^{n}_B(x_1,b_1)\phi_{D}(x_2,b_2) \phi_\pi(x_3)
\nonumber \\
& &\times \alpha_s(t_f)\frac{C_2(t_f)}{N}\exp[-S(t_f)|_{b_3=b_2}]
\nonumber \\
& &\times \Biggl[
x_2  h^{(2)}_f(x_i,b_i)
\Biggr]\;.
\end{eqnarray}
The definitions of the functions, $h_a$, $h_b$ and so on are 
given in \cite{KKLL}. 
Note that the sum of contributions
from $\phi_B$, $\phi^{n+}_B$ and $\phi^{n-}_B$
vanishes if $\phi_B = \phi^{n+}_B = \phi^{n-}_B$.

\end{document}